\documentclass[%
 aip,
 jmp,%
 amsmath,amssymb,
preprint,%
]{revtex4-1}

\usepackage{graphicx}
\usepackage{dcolumn}
\usepackage{bm}

\begin{document}

\def\kket{\rangle \mskip -3mu \rangle}
\def\bbra{\langle \mskip -3mu \langle}

\def\ket{\rangle}
\def\bra{\langle}

\def\pard{\partial}

\def\sinh{{\rm sinh}}
\def\sgn{{\rm sgn}}

\def\alp{\alpha}
\def\del{\delta}
\def\Del{\Delta}
\def\eps{\epsilon}
\def\gam{\gamma}
\def\sig{\sigma}
\def\kap{\kappa}
\def\lam{\lambda}
\def\ome{\omega}
\def\Ome{\Omega}

\def\th{\theta}
\def\vphi{\varphi}

\def\Gam{\Gamma}
\def\Ome{\Omega}

\def\kav{{\bar k}}
\def\vb{{\bar v}}

\def\abf{{\bf a}}
\def\cbf{{\bf c}}
\def\dbf{{\bf d}}
\def\gbf{{\bf g}}
\def\kbf{{\bf k}}
\def\lbf{{\bf l}}
\def\nbf{{\bf n}}
\def\pbf{{\bf p}}
\def\qbf{{\bf q}}
\def\rbf{{\bf r}}
\def\ubf{{\bf u}}
\def\vbf{{\bf v}}
\def\xbf{{\bf x}}
\def\Cbf{{\bf C}}
\def\Dbf{{\bf D}}
\def\Kbf{{\bf K}}
\def\Pbf{{\bf P}}
\def\Qbf{{\bf Q}}

\def\omet{{\tilde \ome}}
\def\gammat{{\tilde \gamma}}
\def\Ft{{\tilde F}}
\def\gt{{\tilde g}}
\def\Ht{{\tilde H}}
\def\tt{{\tilde t}}
\def\Ut{{\tilde U}}
\def\ut{{\tilde u}}
\def\bt{{\tilde b}}
\def\Vt{{\tilde V}}
\def\vt{{\tilde v}}
\def\xt{{\tilde x}}

\def\ph{{\hat p}}

\def\vt{{\tilde v}}
\def\wt{{\tilde w}}
\def\phit{{\tilde \phi}}
\def\rhot{{\tilde \rho}}
\def\Ft{ {\tilde F}}

\def\Cb{{\bar C}}
\def\Nb{{\bar N}}
\def\Ab{{\bar A}}
\def\Bb{{\bar B}}
\def\Db{{\bar D}}
\def\Hb{{\bar H}}
\def\Vb{{\bar V}}
\def\etab{{\bar \eta}}
\def\gb{{\bar g}}
\def\nb{{\bar n}}
\def\bb{{\bar b}}
\def\Pib{{\bar \Pi}}
\def\rhob{{\bar \rho}}
\def\phib{{\bar \phi}}
\def\psib{{\bar \psi}}
\def\omeb{{\bar \ome}}

\def\Sh{{\hat S}}
\def\Wh{{\hat W}}

\def\SS{I}
\def\psiw{{\xi}}
\def\tI{{g}}

\def\Ep#1{Eq.\ (\ref{#1})}
\def\Eqs#1{Eqs.\ (\ref{#1})}
\def\EQN#1{\label{#1}}

\newcommand{\beqa}{\begin{eqnarray}}
\newcommand{\eeqa}{\end{eqnarray}}

\title{Complex spectral analysis and test function spaces}

\author{Sungyun Kim}
\email{rdecay@gmail.com}
 \affiliation{Hoseo University, 165 Sechul Li, Baebang Myun, Asan, Chungnam 336-795, Korea
}%

\date{\today}

\begin{abstract}
 We consider complex eigenstates of unstable Hamiltonian and its physically meaningful regions.
 Starting from a simple model of a discrete state interacting with a continuum via a general potential, we show that its Lippmann-Schwinger
 solution set can be decomposed into a free-field set, a set containing lower half plane pole of Green's function and a set containing upper half pole of Green's function. From here distinctive complex eigenstates corresponding to each pole are constructed. We note that on the real line square integrable functions can be decomposed into Hardy class above and below functions which behave well in their respective complex half planes.
 Test function restriction formulas which remove unphysical growth are given. As a specific example we consider Friedrichs model which solutions and complex eigenstates are known, and compare numerically calculated total time evolution with test function restricted complex eigenstates for various cases. The results shows that test function restricted complex eigenstates capture the essence of decay phenomena quite well.
\end{abstract}

\maketitle

\section{\label{sec:intro}Introduction}

The introduction of complex eigenstate has a long history. Since Gamow proposed complex energy eigenstate to
 describe decay process \cite{Gamow}, it has been widely used for its simplicity and predicting power for decay rate.
 Matthews and  Salam~\cite{Matthews1,Matthews2, Salam}
 developed  the concept of an unstable particle in terms of asymptotic
 states. Nakanishi \cite{Nakanishi} introduced
 complex distributions to
 define a complex eigenstate of the Hamiltonian in Lee's model \cite{Lee}.
  The contour deformation method in momentum space
 is also studied and applied in nuclear physics
 \cite{Brayshaw,Nuttall,Stelbovics}.
Sudarshan, Chiu and Gorini
  \cite{Sudarshan} constructed complex eigenstates  using contour
  deformation in the complex plane.  Bohm and Gadella \cite{Bohm} constructed complex
  eigenvectors using poles of the $S$ matrix and Hardy class test functions (see also \cite{Tasaki}).
 Prigogine and collaborators
   studied extensively the properties of complex spectral representations in
  the Friedrichs model \cite{PPT}, and defined unstable
  states in Liouville space (see \cite{OPP} and references therein).

 But its physical meaning has been in debate. The complex eigenstate itself is not normalized and
 its resemblance to the real wave function
 evolution ceases to exist outside certain spacetime region.
  When the decay occurs it has starting time,
  and the decay products
 propagate according to the causality conditions. Since complex eigenstates' physically meaningful spacetime regions
 is related to its initial conditions or test functions applied to them, it is desirable that complex eigenstates are presented with its meaningful test function criteria.

  Notable researches to this direction have been done by Bohm and Gadella \cite{Bohm},
 who proposed certain class of functions, namely Hardy class functions, should be applied to complex
 eigenstates to obtain physically meaningful results. They also proposed rigged Hilbert spaces to distinguish
 functionals and test functions \cite{BohmQM}. This removed exponential blowup in negative time, but spatial
 exponential growth in complex eigenstate still remained. One of other suggested methods for removing divergences
 is complex scaling method \cite{Moiseyev1} which introduced complex coordinate to diverging regions at large distances.

  In this article we develop a method to find physical meanings of complex eigenstates in all spacetime regions. We start
  from a decay model of one discrete state and continuum states interacting through a general potential and obtain its energy eigenstate set from Lippmann-Schwinger equation solution. These complete eigensets are decomposed into free field part, lower half plane pole part and upper half plane pole part. Two complex eigenstates are constructed from this decomposition, corresponding to upper and lower complex poles. We point out that these two complex eigenstates have their own physically meaningful regions,
  and only parts of test functions are needed for physically meaningful pole contribution.
   For meaningful test function criteria we consider Hardy class function as Bohm and Gadella did\cite{BohmQM} but with improvements and modifications.
  Hardy class function decomposition formulas are given for square integrable functions, and for the decomposed Lippmann-Schwinger solution set the Hardy class function restriction formulas are given for various cases.

   To test the validity of our construction we test our formalism with Friedrichs model for which complete solutions and complex eiggenstates are already known. For a discrete state and field test functions total time evolution and test function restricted complex eigenstates are numerically calculated and compared. It is shown that test function restricted complex eigenstates capture the essence of decay well and demonstrate the physically valid regions of spacetime for corresponding complex eigenstates.

\section{\label{sec:Lippmann} Lippmann-Schwinger equation and S matrix}
We start with standard quantum scattering setting. Consider a simple Hamiltonian $H_0$ with a discrete state and energy continuum.
 \beqa
  H_0 = \ome_1 |1\ket\bra 1| + \int_0^{\infty}d\ome\, \ome|\ome\ket \bra \ome|   \EQN{1H0}
 \eeqa
 Without interaction between the discrete state and continuous states, $|1\ket$ state is just a bound state. The continuous energy spectrum spans from 0 to $\infty$ (bounded from below), preventing negative infinite energy.
 In this setup $H_0$ eigenstates are orthonormal and complete, i.e.
 \beqa
   & &\bra 1|1\ket=1,\,\,\, \bra \ome'|\ome\ket =\delta(\ome'-\ome),\,\,\,
   \bra 1|\ome\ket =\bra \ome |1\ket =0.  \\
    & & |1\ket\bra 1| + \int_0^{\infty}d\ome\, |\ome\ket \bra \ome| =1.
 \eeqa
   When there is interaction between the discrete state and continuum, situation changes. Let us write
 the interaction term $V$. The total Hamiltonian $H$ is written as
 \beqa
  H= H_0 + V \EQN{3H0V}
  \eeqa
  and $|1\ket$, $|\ome\ket$ are no more energy eigenstates of $H$.
 When the discrete energy is within the continuous energy spectrum ($\ome_1 >0$ in our case) and total energy is finite, the discrete energy state may decay into the continuum.
   Known methods to analyze \Ep{3H0V} are to find total eigenstates of $H$, or to find a governing equation for total eigenstates. In general exact analytic solution for total eigenstates is known only for special cases of $V$, but the governing equation for total eigenstates are well known. The governing equation is known as Lippmann-Schwinger equation and one of most used equations in collisions and quantum scattering. With our eigenstates of $H_0$, $|1\ket$ and $\ome\ket$, the energy eigenstate $|\Psi^{\pm}_{\ome}\ket$ of total Hamiltonian $H$ can be written as \cite{BohmQM}
   \beqa
   & & |\Psi^{+}_{\ome}\ket = |\ome\ket + \frac{1}{\ome-H_0 + i\eps}V |\Psi^{+}_{\ome}\ket= |\ome\ket + \frac{1}{\ome-H + i\eps}V |{\ome}\ket, \EQN{4Fp} \\
   & & |\Psi^{-}_{\ome}\ket = |\ome\ket + \frac{1}{\ome-H_0 - i\eps}V |\Psi^{-}_{\ome}\ket =|\ome\ket + \frac{1}{\ome-H - i\eps}V |{\ome}\ket \EQN{4Fpm}
    \eeqa
    where $\eps$ is a positive infinitesimal.
    In case that the discrete state decays into continuum, the total energy eigenstates set does not contain discrete energy state. With our specific form of $H_0$ in \Ep{1H0}, \Ep{4Fpm} can be written as
 \beqa
   & &|\Psi^{+}_{\ome}\ket = |\ome\ket + \frac{1}{\ome-H_0 + i\eps}V |\Psi^{+}_{\ome}\ket \nonumber \\
   & &= |\ome\ket + (|1\ket\bra 1| + \int_0^{\infty}d\ome' |\ome'\ket \bra \ome'| )\,\,\frac{1}{\ome-H_0 + i\eps}V |\Psi^{+}_{\ome}\ket \nonumber \\
   & &= |\ome\ket + |1\ket\bra 1|\Psi_\ome^+\ket + \int_0^{\infty}d\ome' \frac{|\ome'\ket\bra\ome'|V|\Psi_\ome^+\ket}{\ome-\ome'+ i \eps}, \EQN{5F+} \\
   & &|\Psi^{-}_{\ome}\ket = |\ome\ket + |1\ket\bra 1|\Psi_\ome^-\ket + \int_0^{\infty}d\ome' \frac{|\ome'\ket\bra\ome'|V|\Psi_\ome^-\ket}{\ome-\ome'- i \eps}.
 \eeqa
where we used
 \beqa
 \bra 1|\Psi_\ome^+\ket = \frac{\bra 1|V|\Psi_\ome^+\ket}{\ome-\ome_1+i\eps}
 \eeqa
  by applying $\bra 1|$  on \Ep{4Fpm}.

  It  is known \cite{EnssV} that both $|\Psi^{\pm}_{\ome}\ket$ states form complete and orthogonal sets, namely
    \beqa
     \bra \Psi_{\ome'}^{+}|\Psi_{\ome}^+\ket =  \bra \Psi_{\ome'}^{-}|\Psi_{\ome}^-\ket = \delta (\ome'-\ome) \\
      \int_0^{\infty} |\Psi_{\ome}^+\ket\bra \Psi_{\ome}^+| = \int_0^{\infty} |\Psi_{\ome}^-\ket\bra \Psi_{\ome}^-| = 1.
    \eeqa

   In the discussion of scattering and decay, the wave operators, scattering operator, scattering matrix and its poles are frequently mentioned. The M{\o}ller wave operators $\Omega^{\pm}$ satisfy the relations
 \beqa
  \Omega^+ |\ome\ket = |\Psi_\ome^+\ket, \,\,\,  \Omega^- |\ome\ket = |\Psi_\ome^-\ket,
 \eeqa
  and the scattering operator $\bold{S}$ is defined as
  \beqa
   \bold{S}= \Omega^{-\dagger} \Omega^+
  \eeqa
 and the scattering matrix ($S$ matrix) is obtained by
  \beqa
   S_{\ome,\ome'} =\bra \ome'| \bold{S} |\ome\ket = \bra \Psi^{-}_{\ome}| \Psi_{\ome}^+ \ket.
   \eeqa
  Since
 \beqa
  |\Psi_\ome^+\ket = |\Psi_\ome^-\ket - 2 \pi i \delta (\ome-H)  V|\ome\ket,
\eeqa
$S$ matrix has the form
 \beqa
  S_{\ome' \ome} =  \bra \Psi_{\ome'}^-|\Psi^+_{\ome}\ket
  = \delta(\ome-\ome') (1-2\pi i \bra \Psi_{\ome'}^-| V|\ome\ket)
 \eeqa
 and if $\bra \Psi_{\ome'}^-|V|\ome\ket$ has no poles for $\ome=\ome'$ one can write
 \beqa
  & &S_{\ome' \ome} = S(\ome) \delta (\ome-\ome') \\
  & &S(\ome) =  1-2\pi i \bra \Psi_{\ome}^-| V|\ome\ket =1- 2\pi i \bra \ome| V| \Psi_{\ome}^+ \ket. \EQN{18SFp} \\
  & &|\Psi^+_{\ome}\ket = S(\ome) |\Psi^{-}_{\ome}\ket. \EQN{17FS}
  \eeqa
   From the unitarity of $\bold{S}$ one can also see
  \beqa
   S^{-1}(\ome) = 1+2\pi i \bra \Psi_{\ome}^+| V|\ome\ket=1+ 2\pi i \bra \ome| V| \Psi_{\ome}^- \ket. \EQN{19Sin}
  \eeqa
  The poles of $S$ matrix are related to the resonance and decay states. If we assume that the potential can be
  analytically continued to the lower half plane, S matrix is known to have poles in the lower half plane \cite{JTaylor}.
  Also, the complex conjugate of the $S$ matrix poles are the zeros of $S$ matrix, or poles of $S^{-1}$ matrix.

  In our case of $\ome_1 >0$, which means the discrete energy is inside the continuous spectrum, $S$ matrix pole just below the real axis corresponds to the decay state of $|1\ket$. Or the upper half plane pole of $S^{-1}$ matrix corresponds to the growing state of $|1\ket$.

   But one should note that not every part of total Hamiltonian eigenstate is related to the scattering or decay state.  As one see in \Ep{4Fp}, $|\Psi^{+}_\ome\ket$ consists of free field term plus scattering term due to the interaction. $S$ matrix pole are related to scattering term, and depending on the initial conditions scattering or decay effects might be small or large. It is desirable that terms related to the S matrix poles are separated from free field term if one wants to see the scattering or decaying effect clearly. 

    We show that there is a way to separate free field term, decay term and growing term in complete eigenstate set of our system. We can write, after some manipulation (see appendix [\ref{App1}]),
   \beqa
    & &|\Psi^+_\ome\ket\bra \Psi^+_\ome|  \nonumber \\
    & &=    \bigg( |\ome\ket + |1\ket\bra 1|\Psi_\ome^+\ket + \int_0^{\infty}d\ome' \frac{|\ome'\ket\bra\ome'|V|\Psi_\ome^+\ket}{\ome-\ome'+ i \eps} \bigg) \nonumber \\
    & &\times \bigg( \bra \ome | +\bra \Psi_\ome^+|1\ket\bra 1| + \int_0^{\infty}d\ome' \frac{\bra \Psi_\ome^+|V|\ome'\ket \bra \ome'| }{\ome-\ome'-i\eps} \bigg) \nonumber \\
  & &= |\ome\ket\bra\ome| \nonumber \\
  & &- \bra 1 |\Psi_\ome^+\ket \frac{\bra 1|\Psi_\ome^+\ket}{2 \pi i \bra \ome|V|\Psi_\ome^+\ket } \bigg( |1\ket + \int_0^{\infty}d\ome' \,\frac{|\ome'\ket}{\ome-\ome'+i\eps}   \frac{\bra \ome'|V|\Psi_\ome^+\ket}{\bra 1 |\Psi_\ome^+\ket} \bigg) \nonumber \\
  & &\times \bigg( \bra 1| + \int_0^{\infty}d\ome' \, \frac{\bra \ome'|} {\ome-\ome'+i\eps} \frac{\bra \Psi_\ome^+ |V|\ome'\ket}{\bra \Psi_\ome^+ |1 \ket} \bigg) \nonumber \\
  & &+ \bra \Psi_\ome^+|1\ket\frac{\bra 1|\Psi_\ome^+\ket}{2\pi i\bra \ome|V|\Psi_\ome^+\ket } \bigg( |1\ket + \int_0^{\infty}d\ome' \,\frac{|\ome'\ket}{\ome-\ome'-i\eps} \frac{\bra \ome'|V|\Psi_\ome^+\ket}{\bra 1 |\Psi_\ome^+\ket} \bigg) \nonumber \\
   & &\times \bigg( \bra 1| +\int_0^{\infty}d\ome' \, \frac{\bra \ome'|} {\ome-\ome'-i\eps} \frac{\bra \Psi_\ome^+ |V|\ome'\ket}{\bra \Psi_\ome^+ |1 \ket} \bigg) \nonumber \\
  & &=|\ome\ket\bra\ome| +  A_\ome \bigg( |1\ket + \int_0^{\infty}d\ome' \frac{|\ome'\ket B_{\ome'\ome}}{\ome-\ome'+i\eps}\bigg) \bigg( \bra 1| + \int_0^{\infty}d\ome' \frac{\bra \ome'|  B_{\ome'\ome}^{c.c}}{\ome-\ome'+i\eps}\bigg) \nonumber \\
  & &+A_\ome^{c.c} \bigg( |1\ket + \int_0^{\infty}d\ome' \frac{|\ome'\ket B_{\ome'\ome}}{\ome-\ome'-i\eps}\bigg) \bigg( \bra 1| + \int_0^{\infty}d\ome' \frac{\bra \ome'|  B_{\ome'\ome}^{c.c}}{\ome-\ome'-i\eps}\bigg)
   \EQN{20Sep}
  \eeqa
   where
  \beqa
   A_\ome = - \bra 1 |\Psi_\ome^+\ket \frac{\bra 1|\Psi_\ome^+\ket}{2 \pi i \bra \ome|V|\Psi_\ome^+\ket },
   \,\,\, B_{\ome' \ome}= \frac{\bra \ome'|V|\Psi_\ome^+\ket}{\bra 1|\Psi_\ome^+  \ket}
   \eeqa
 \Ep{20Sep} has very claer separations of terms which have distinctive meanings. First term in \Ep{20Sep} is free field eigenstate expansion.  Second term is related to the $S(\ome)$ poles which describe decaying state. Third term is related to the $S^{-1}(\ome)$ poles which describe growing states.

  With this separation we can see the effect of complex poles more clearly and how each terms behave in their dominant spacetime regions. Next section we apply a specific model to \Ep{20Sep} which allows us a direct theoretical
  and numerical calculations.

\section{Pole extraction and complex eigenstates  }

 In this section we derive complex eigenstates and
 study its test function conditions. For this purpose construction
 process of complex eigenstate is examined first.

 In many cases the complex eigenstates are constructed from the pole of S-matrix. Generally S-matrix might have infinite
 number of poles, so to get the physically meaningful complex eigenstate one should first identify the physically
 meaningful pole. In our case we are investigating the decay phenomena of discrete state $|1\ket$, which results from
 the interaction between discrete state and field states. When the interaction is small, the decay is slow and distinctive. This corresponds to the fact that there is a pole in the lower half plane which has small imaginary part and real part close to the discrete state eigenenergy $\ome_1$. The complex eigenstate related to the decay of $|1\ket$ is constructed with this pole.

 From \Ep{17FS}, we can see that
  \beqa
   S(\ome) = \frac{\bra 1|\Psi_\ome^+\ket}{\bra 1| \Psi_\ome^-\ket}= \frac{\bra 1|\Psi_\ome^+\ket}{\bra \Psi_\ome^+|1\ket} \EQN{23S1}
  \eeqa
 in our model. From \Ep{4Fp}, we can write
 \beqa
 \bra 1|\Psi_\ome^+\ket = \bra 1|  \frac{1}{\ome- H + i\eps}V |{\ome}\ket = \int_0^{\infty} d\ome' \frac{\bra 1 | \Psi_{\ome'}^+\ket \bra \Psi_{\ome'}^+|V|\ome\ket }{\ome-\ome'+i\eps} \EQN{24_1w}
 \eeqa
  and we see that when $\ome$ in \Ep{24_1w} changes to complex number in the upper half plane the integration is well defined whenever the integral for real $\ome$ is defined, except possible singularities due to $\bra \Psi_{\ome'}^+|V|\ome\ket$.

  The poles
  from the Green's function $1/(\ome^+ -H)$  appear only in the lower half plane ( we denote $\ome^+$ and $\ome^-$ as complex energy obtained by analytic continuation from upper or lower half plane)
   and these
  poles also correspond to the complex eigenvalues of total Hamiltonian.

 Similarly, $\bra \Psi_\ome^+|1\ket$ has no poles in the lower half plane except $\bra \ome|V| \Psi_{\ome'}^+\ket$ poles and the Green's function $1/(\ome^- -H)$ poles appear in the upper half plane.

    From the above discussion we see that $S(\ome)$ in \Ep{23S1} has poles in the lower half plane, which are coming from $1/(\ome^+ -H)$, and zeroes in the upper half plane which comes from $1/(\ome^- -H)$ poles.  The poles coming from potential term $\bra \Psi_{\ome'}^+|V|\ome\ket$ cancel out.

  With the knowledge that the poles of $S(\ome)$ correspond to the complex eigenvalues of Hamiltonian, let us see \Ep{20Sep}. We examine the term
   \beqa
  & &- \bra 1 |\Psi_\ome^+\ket \frac{\bra 1|\Psi_\ome^+\ket}{2 \pi i \bra \ome|V|\Psi_\ome^+\ket } \bigg( |1\ket + \int_0^{\infty}d\ome' \,\frac{|\ome'\ket}{\ome-\ome'+i\eps}   \frac{\bra \ome'|V|\Psi_\ome^+\ket}{\bra 1 |\Psi_\ome^+\ket} \bigg) \nonumber \\
  & &\times \bigg( \bra 1| + \int_0^{\infty}d\ome' \, \frac{\bra \ome'|} {\ome-\ome'+i\eps} \frac{\bra \Psi_\ome^+ |V|\ome'\ket}{\bra \Psi_\ome^+ |1 \ket} \bigg). \EQN{40secd}
    \eeqa
   and its coefficient
\beqa
 - \bra 1 |\Psi_\ome^+\ket \frac{\bra 1|\Psi_\ome^+\ket}{2 \pi i \bra \ome|V|\Psi_\ome^+\ket } . \EQN{41coef}
 \eeqa
  From \Ep{18SFp}
 we see that $\bra \ome|V|\Psi_\ome^+\ket$ has the same  poles as $S(\ome)$, so $\bra 1|\Psi_\ome^+\ket / \bra \ome|V|\Psi_\ome^+\ket $ does not have $1/(\ome^+ -H)$ poles.  The $1/(\ome^+ -H)$ lower half plane poles in the term (\ref{41coef}) comes from $ \bra 1 |\Psi_\ome^+\ket $. Inside the bracket the term
  \beqa
  \int_0^{\infty}d\ome' \,\frac{|\ome'\ket}{\ome-\ome'+i\eps}  \frac{\bra \ome'|V|\Psi_\ome^+\ket}{\bra 1 |\Psi_\ome^+\ket}
  \eeqa
  has $|\Psi_\ome^+\ket$ in both numerator and denominator, so the possible  poles from $1/(\ome^+ -H)$
  cancel out. As a whole the term (\ref{40secd}) has the lower half plane poles from $1/(\ome^+ -H) $
  and no poles from $1/(\ome^- -H) $.

  Following similar reasoning, we can show that in \Ep{20Sep},
  \beqa
 & & \bra \Psi_\ome^+|1\ket\frac{\bra 1|\Psi_\ome^+\ket}{2\pi i\bra \ome|V|\Psi_\ome^+\ket } \bigg( |1\ket + \int_0^{\infty}d\ome' \,\frac{|\ome'\ket}{\ome-\ome'-i\eps} \frac{\bra \ome'|V|\Psi_\ome^+\ket}{\bra 1 |\Psi_\ome^+\ket} \bigg) \nonumber \\
   & &\times \bigg( \bra 1| +\int_0^{\infty}d\ome' \, \frac{\bra \ome'|} {\ome-\ome'-i\eps} \frac{\bra \Psi_\ome^+ |V|\ome'\ket}{\bra \Psi_\ome^+ |1 \ket} \bigg)
   \eeqa
  has upper half plane poles from  $1/(\ome^- -H)$ and no poles from $1/(\ome^+ -H) $.

  These results give very nice separations with respect to complex energies. When we want to construct complex eigenstate with eigenenergies in the upper half plane, we need to consider only the second term in \Ep{20Sep}. Other terms do not contribute. Likewise, when we construct complex eigenstate in the lower half plane, we need to consider only the third term in \Ep{20Sep}.

  The complex eigenstates can be constructed by deforming the real line integration into the pole encircling contour plus the rest. Let us call the lower half plane pole which is close to $\ome_1$ as $z$. From \Ep{20Sep} we can write
   \beqa
   & &\int_0^{\infty} d\ome |\Psi_\ome^+\ket \bra \Psi_\ome^+| = \int_0^{\infty}d\ome \,\,|\ome\ket\bra\ome| \nonumber \\
   & &+ \bigg( \int_{C_{z}} d\ome + \int_{R^+ - C_{z}} d\ome \bigg)  A_\ome \bigg( |1\ket + \int_0^{\infty}d\ome' \frac{|\ome'\ket B_{\ome'\ome}}{\ome-\ome'+i\eps}\bigg) \bigg( \bra 1| + \int_0^{\infty}d\ome' \frac{\bra \ome'|  B_{\ome'\ome}^{c}}{\ome-\ome'+i\eps}\bigg) \nonumber \\
   & &+ \bigg( \int_{C_{z^{c.c}}} d\ome + \int_{R^+ - C_{z^{c.c}}} d\ome \bigg)  A_\ome^{c} \bigg( |1\ket + \int_0^{\infty}d\ome' \frac{|\ome'\ket B_{\ome'\ome}}{\ome-\ome'-i\eps}\bigg) \bigg( \bra 1| + \int_0^{\infty}d\ome' \frac{\bra \ome'|  B_{\ome'\ome}^{c}}{\ome-\ome'-i\eps}\bigg) \nonumber \\
  & &= \int_0^{\infty}d\ome \,\,|\ome\ket\bra\ome| + |\phi_{z}\ket\bra \phit_{z}| +
  |\phi_{z^{c.c}}\ket\bra \phit_{z^{c.c}}| + (\mbox{rest}). \EQN{29comp}
   \eeqa
   In \Ep{29comp}, $ \int_{C_{z}} d\ome$ is the contour integral which encircles $z$ in clockwise direction, and $ \int_{C_{z^{c.c}}} d\ome$ is the contour integral which encircles $z^{c.c}$ in counterclockwise direction. $\int_{R^+ - C_{z}} d\ome$ and $\int_{R^+ - C_{z^{c.c}}} d\ome$ are the nonnegative real line minus $z$ or $z^{c.c}$ pole encircling contours, respectively. Also we used new notations $A_\ome^{c}$ and $B_{\ome'\ome}^{c}$, since the complex conjugate relations might hold only for real $\ome$.
   $A_\ome^{c}$ and $B_{\ome'\ome}^{c}$ of complex variables are defined to mean that they are analytically continued from  $A_\ome^{c.c}$ and $B_{\ome'\ome}^{c.c}$ of real variables.

$|\phi_{z}\ket\bra \phit_{z}|$ and $ |\phi_{z}\ket\bra \phit_{z}|$   are defined as
      \beqa
   & & |\phi_{z}\ket\bra \phit_{z}| \nonumber \\
   & &=  \int_{C_{z}} d\ome  A_\ome \bigg( |1\ket + \int_0^{\infty}d\ome' \frac{|\ome'\ket B_{\ome'\ome}}{\ome-\ome'+i\eps}\bigg) \bigg( \bra 1| + \int_0^{\infty}d\ome' \frac{\bra \ome'|  B_{\ome'\ome}^{c}}{\ome-\ome'+i\eps}\bigg) \nonumber \\
  & &= N \bigg( |1\ket + \int_0^{\infty}d\ome' \,\frac{|\ome'\ket  B_{\ome' z}}{{z}-\ome'} - 2\pi i B_{z z} |z\ket \bigg) \nonumber \\
   & &\times \bigg( \bra 1| + \int_0^{\infty}d\ome' \, \frac{B_{\ome' z}^{c}  \bra \ome'|} {{z}-\ome'} - 2 \pi i B_{z z}^{c}
  \bra z | \bigg), \EQN{28Psiz}
  \eeqa
  \beqa
  & & N = \lim_{\ome \rightarrow z} (-2\pi i) (\ome-z) A_\ome \EQN{47Ndef}
  \eeqa
   \beqa
   & &|\phi_{z^{c.c}}\ket\bra \phit_{z^{c.c}}| \nonumber \\
   & & =
\int_{C_{z^{c.c}}} d\ome  A_\ome^{c} \bigg( |1\ket + \int_0^{\infty}d\ome' \frac{|\ome'\ket B_{\ome'\ome}}{\ome-\ome'-i\eps}\bigg) \bigg( \bra 1| + \int_0^{\infty}d\ome' \frac{\bra \ome'|  B_{\ome'\ome}^{c}}{\ome-\ome'-i\eps}\bigg)  \nonumber \\
   & &=  N^{c.c} \bigg( |1\ket + \int_0^{\infty}d\ome' \,\frac{|\ome'\ket B_{\ome' z^{c.c}}}{{z^{c.c}}-\ome'} + 2\pi i B_{\ome' z^{c.c}} |z^{c.c}\ket \bigg) \nonumber \\
   & &\times \bigg( \bra 1| + \int_0^{\infty}d\ome' \, \frac{B_{\ome' z^{c.c}}^{c} \bra \ome'|} {{z^{c.c}}-\ome'} + 2 \pi i B_{\ome' z^{c.c}}^{c} \bra z^{c.c} | \bigg), \EQN{47z1cc}
   \eeqa
   and $(\mbox{rest})$ term in \Ep{29comp} are   $\int_{R^+ - C_{z}} d\ome$ and $\int_{R^+ - C_{z^{c.c}}} d\ome$ integrations. Now the question arises, whether the quantities like $|\Psi_{z}^+\ket$, $|z\ket$ can be well defined. The answers depend on the properties of system as well as initial and final conditions. Terms like
   $\bra z|V| \Psi_{z}^+\ket$ and $\bra 1| \Psi_{z}^+\ket$ can be analytically continued from real line to $z$ depending on the form of the potential $V$ \cite{JTaylor}. Their existence depends on the system structure. On the contrary, definability of terms like $\bra f|z\ket$ and  $\bra z^{c.c}|g\ket$ depend on the analyticity of test functions $\bra f|$ and $|g\ket$ which act on the $|z\ket$ and $\bra z^{c.c}|$. These test functions are not related to the system structure but the initial and final conditions which are usually square integrable
   but not necessarily analytic. Even when the potential $V$ behaves well analytically and $S(\ome)$ is well defined, the complex eigenstates have meanings only for certain types of test functions.

For the discussion of suitable choice of test functions, let us consider a simple integral which contains a pole and its integration is over nonnegative real axis. Suppose that we have an integration given by
   \beqa
   F(y) = \int_0^{\infty}d\ome \frac{e^{i \ome |x|}+ e^{-i \ome |x|} }{\ome-y}. \EQN{31Fz}
    \eeqa
    When $Im[y]$ is very small and $Re[y] |x|>0$ is not very close to zero , the integral is dominated by near $ Re[y]$ integration. In this case one might add negative real axis integration and approximate
  \beqa
  \int_0^{\infty}d\ome \frac{e^{i \ome |x|}+ e^{-i \ome |x|}}{\ome-y} \approx \int_{-\infty}^{\infty}d\ome \frac{e^{i \ome |x|}+ e^{-i \ome |x| }}{\ome-z} = -2\pi i e^{-i y |x|}.
  \eeqa
   In this approximation only part of integrating function $e^{-i\ome |x|}$ contributes and other part vanishes.
   We easily see why the result of this integration is different from pole encircling contour integration of \Ep{29comp}. $e^{i\ome |x|}$ contribution does not exist in the whole real line integration but is included
   in pole encircling integration.

   From this simple example we can find hints for constructing physically meaningful complex eigenstates. First
   one might suggest integrating over the whole real line instead of positive real line for complex eigenstates. But directly extending integrations to the whole line can produce many unnecessary terms if the test function contains other poles than $z$. The $z$ pole encircling contour method has advantage of taking only $z$ pole contributions, so we have to consider a way to take only $z$ residue part while removing unphysical growth.

    We see that this might be possible if we take only part of test functions in the pole encircling integration.
    This part of test functions should be analytic and should not grow for parameters like $t$ and $x$ if the original real
  line integration does not grow on those parameters.
   Fortunately
   there exist class of functions that make square integrable functions into sum of analytic functions of corresponding domains. They are called Hardy class functions and it is shown in later section
  \beqa
  \int_{-\infty}^{\infty}d\ome \frac{f(\ome)}{\ome-y}
  = -2 \pi i [f(y)]^- \mbox{ for $Im[y]<0$}
  \eeqa
  where $[f(y)]^-$ is Hardy class below part of $f(y)$.

   A very nice property of hardy class decomposition is that this decomposition fixes both exponential grow problem and analyticity problem of test functions simultaneously.

 Next section we examine the definitions and properties of Hardy class functions.

 \section{ Hardy class functions on the real line}

   In this section Hardy class functions are defined and properties are summarized \cite{Duren, BohmQM}. Hardy class function
   decomposition formula is presented and important points are noted.

 A complex function $G(E)$ on the real line is a Hardy class function from above (below) if
  $G(E)$ is the boundary value of an analytic function G($\ome$) in the upper (lower) half plane and
     \beqa
      \int_{-\infty}^{\infty}dE | G(\ome)|^p < \infty
      \eeqa
  for all $\ome$ in the upper (lower) half plane. Here we consider $p=2$ cases since in quantum mechanics we deal with
  square integrable functions. The spaces of above or below Hardy class functions are denoted as $H_+^2$ or $H_-^2$.

 Hardy class functions has important properties that can be quite useful for our constructions.

  (1) If $G(\ome)$ is in $H_\pm^2$ then $G(\ome)$ on the real axis is uniquely determined by its values on the positive real
  axis.

  (2) If $G(\ome)$ is in $H_+^2$ then
   \beqa
    \frac{1}{2\pi i} \int_{-\infty}^{\infty}d\ome \frac{G(\ome)}{\ome-y} =\left\{  \begin{array}{ll}  G(z)
     & \mbox{for $Im[y]>0$} \\
     0 & \mbox{for $Im[y]<0$} \\
     \end{array} \right. ,
   \eeqa
   If $G(\ome)$ is in $H_-^2$ then for all $Im[z]<0$
\beqa
    \frac{-1}{2\pi i} \int_{-\infty}^{\infty}d\ome \frac{G(\ome)}{\ome-y} =\left\{  \begin{array}{ll}  0
     & \mbox{for $Im[y]>0$} \\
     G(y) & \mbox{for $Im[y]<0$} \\
     \end{array} \right.
     \eeqa

    (3) If $G_{\pm}(\ome)$ is in $H_\pm^2$, then the Fourier transform
    \beqa
     \hat G_{\pm}(t) = \frac{1}{\sqrt{2\pi}} \int_{-\infty}^{\infty}d\ome e^{-i\ome t} G(\ome)
       \eeqa
      has the property
       \beqa
      & & \hat G_+(t) = 0 \,\,\,\mbox{for $t<0$}  \\
       & &\hat G_-(t) = 0 \,\,\,\mbox{for $t>0$}
       \eeqa
       (Paley-Wiener Theorem)

       (4) If $G(\ome)$ is in $H_+^2$, its complex conjugate is in $H_-^2$ and vice versa.
 \newline

      From the above properties one can observe that there is a quite simple way to decompose square integrable functions on the real line into
       Hardy class functions. One can write
       \beqa
       f(\ome) = \frac{-1}{2\pi i} \int_{-\infty}^{\infty} d\ome' \frac{f(\ome')}{\ome-\ome'+i\eps} +
       \frac{1}{2\pi i} \int_{-\infty}^{\infty} d\ome' \frac{f(\ome')}{\ome-\ome'-i\eps}
       = \int_{-\infty}^{\infty}d\ome \delta(\ome-\ome')f(\ome')
       \eeqa
       for the positive infinitesimal $\eps$. For square integrable functions, this relation holds if $f(\ome)$ is
       continuous (even if $f(\ome)$ is not continuous
         it might be applicable in physical situations provided discontinuous points set is negligible). When the above relation can be defined, one can make Hardy class decomposition
       on the real line as
\beqa
       f(\ome) = f(\ome)^{+}+ f(\ome)^{-} \EQN{47Hd}
       \eeqa
       with
       \beqa
    f(\ome)^{+} = \frac{-1}{2\pi i}\int_{-\infty}^{\infty} d\ome' \frac{f(\ome')}{\ome-\ome'+i\eps},\,\,\,
     f(\ome)^{-}=  \frac{1}{2\pi i} \int_{-\infty}^{\infty} d\ome' \frac{f(\ome')}{\ome-\ome'-i\eps}. \EQN{44fH-}
       \eeqa

 Usefulness of the decomposition formula \Ep{47Hd} becomes apparent when we consider the behavior of decomposed functions in complex plane. For the variables of finite imaginary part we have
    \beqa
        & &f(y)^{+} = \frac{-1}{2\pi i}\int_{-\infty}^{\infty} d\ome' \frac{f(\ome')}{y-\ome'}
        \;\;\; \mbox{ if $Im[y] >0$,} \EQN{50Hz} \\
          & &f(y)^{-} = \frac{1}{2\pi i}\int_{-\infty}^{\infty} d\ome' \frac{f(\ome')}{y-\ome'}
        \;\;\; \mbox{ if $Im[y] <0$} \EQN{51Hz}
        \eeqa
        and the integrals in \Ep{50Hz} and \Ep{51Hz} converge and are well defined for all $n$ time differentiations with respect to $y$. So the decomposition \Ep{47Hd} expresses
 a square integrable function as
       sum of real line boundary values of two analytic functions, in upper and lower half planes respectively. Square integrable functions need not be differentiable or analytic, but each $H_+$ and $H_-$ functions are analytic and infinitely differentiable in
       their respective domains. Thus this decomposition makes complex analysis possible.

        One can also define Hardy class operators $[ \,\,]^{\pm}$ using \Ep{44fH-} by defining
        \beqa
       & &f(\ome)^{+}= [f(\ome)]^{+} \equiv \frac{-1}{2\pi i}\int_{-\infty}^{\infty} d\ome' \frac{f(\ome')}{\ome-\ome'+i\eps}
       \\
       & &f(\ome)^{-}= [f(\ome)]^{-} \equiv \frac{1}{2\pi i}\int_{-\infty}^{\infty} d\ome' \frac{f(\ome')}{\ome-\ome'-i\eps}.
       \eeqa
      They are linear and orthogonal as
\beqa
       & &[a f(\ome) +b g(\ome)]^{\pm} = a[f(\ome)]^{\pm}+ b[g(\ome)]^{\pm} \\
       & & [f(\ome)^{+}]^{+} = [f(\ome)]^{+},\;\;\; [f(\ome)^{-}]^{-} = [f(\ome)]^{-} \\
       & & [f(\ome)^{-}]^{+} = [f(\ome)^{+}]^{-}=0
       \eeqa

       Hardy class decompositions are frequently used in physics though its name is not often mentioned. When diverging
       denominator such as $1/(E-H)$ is regularized as $1/(E-H\pm i \eps)$, it means this distribution takes Hardy class
       above (below) part of test functions with respect to the variable not integrated, as seen in  \Ep{44fH-}. Calculations of retarded and advanced Green's functions
       and Lippmann-Schwinger equations are well known examples.

       The Hardy class function above (below) behaves very nicely in the upper (lower) half plane. It is analytic and monotonically
       decreases as the absolute value of imaginary part of argument increases. This monotonic decreasing property can be associated
       with physical properties of test functions, like considerations of boundary conditions in
        retarded and advanced Green's function derivation. In our decay problem, test function conditions can be associated with Hardy class conditions. Usually test functions are related to physical initial and final
        states and square integrable (or delta-normalized), but in general these test functions cannot be applied
        to complex eigenstates defined in \Ep{28Psiz}. Square integrable functions are generally not analytic so they
        cannot be analytically continued to lower half plane. Also real line integration is close to the pole enclosing contour integration only when the functions can be continued to the lower half plane and remaining contour integral besides the pole encircling contour is very small. Decomposing square integrable function into Hardy class functions
        and taking only lower part solves these problems, thus suggesting a good way to obtain physically meaningful
        regions of complex eigenstates as shown in next section.

 \section{Hardy class test functions on complex eigenstates} \label{sec:Htest}
       In this section we
       apply Hardy class test functions to the complex eigenstates of
       our model and examine its physical properties. To this end, we first write general time evolved transition
       $\bra f|e^{-iHt}|g\ket$ expanded by real energy eigenstates. Then the test function parts are decomposed into
       Hardy class below and above and finally the pole residue $z$ (or $z^{c.c}$) is taken. In decay problem
       the time evolution of discrete state and field states are of great interests, so their transition probabilities
       are analyzed.

        $\bra f|e^{-iHt}|g\ket$ expanded by real energy eigenstates is written as (see \Ep{20Sep})
        \beqa
        & &\bra f|e^{-iHt}|g\ket =\int_0^{\infty}d\ome \bra f| e^{-i\ome t}|\Psi_\ome^+\ket \bra \Psi_\ome^+|g\ket \nonumber \\
        & &= \int_0^{\infty} d\ome \bra f|\ome\ket e^{-i\ome t}\bra \ome |g\ket  \nonumber \\
        & &+ \int_{C_{z}}d\ome A_\ome \bigg( |1\ket + \int_0^{\infty}d\ome' \frac{|\ome'\ket B_{\ome'\ome}}{\ome-\ome'+i\eps}\bigg) e^{-i\ome t}\bigg( \bra 1| + \int_0^{\infty}d\ome' \frac{\bra \ome'|  B_{\ome'\ome}^{c}}{\ome-\ome'+i\eps}\bigg) \nonumber \\
  & &+\int_{C_{z^{c.c}}} d\ome   A_\ome^{c} \bigg( |1\ket + \int_0^{\infty}d\ome' \frac{|\ome'\ket B_{\ome'\ome}}{\ome-\ome'-i\eps}\bigg) e^{-i\ome t}\bigg( \bra 1| + \int_0^{\infty}d\ome' \frac{\bra \ome'|  B_{\ome'\ome}^{c}}{\ome-\ome'-i\eps}\bigg)  \nonumber \\
   & &+ (\mbox{remaining terms}) \EQN{55fg}
        \eeqa
In last expression of \Ep{55fg} free field parts, lower half plane Green's function poles and upper half plane Green's function poles are separated. The remaining term contribution in \Ep{55fg} is small compared to the pole contributions when poles are close to the real axis and the functions analytically continued to the complex planes do not grow. When the test functions are restricted there is no growth and remaining terms are small. 

 Let us consider the second term in \Ep{55fg}
  \beqa
  & &\bra f|\phi_{z}\ket e^{-i z t} \bra \phit_{z}|g\ket  \nonumber \\
 & & = \int_{C_{z}}d\ome A_\ome \bigg( \bra f|1\ket + \int_0^{\infty}d\ome' \frac{\bra f|\ome'\ket B_{\ome'\ome}}{\ome-\ome'+i\eps}\bigg) e^{-i\ome t}\bigg( \bra 1|g\ket + \int_0^{\infty}d\ome' \frac{\bra \ome'| g\ket B_{\ome'\ome}^{c}}{\ome-\ome'+i\eps}\bigg). \EQN{53z1}
   \eeqa
   This term $ \bra f|\phi_{z}\ket e^{-i z t} \bra \phit_{z}|g\ket$ is diverging for $t<0$, and spatial divergence might appear if $\bra f|$ and $|g\ket$ contain spatial field components. We want to remove the diverging effects and obtain only physically meaningful results, while preserving the complex eigenstate structure as much as possible. To this end we
   restrict test functions $\bra f|$, $|g\ket$, $e^{-i\ome t}$ while not touching the system dependent parts.

For complex eigenstates $| \phi_{z}\ket  \bra \phit_{z}|$ with eigenvalue $z$ in the lower half plane we restrict its test functions as Hardy class below parts ($H_-$ parts). When the test function parts are alone $H_-$ part separation is not difficult, but some parts of test functions are multiplied with interaction potential and inside the integration so they need careful considerations.

       Before we proceed, it is convenient to define a notation for taking Hardy class parts only for test functions. Let us write $\big ( \big)^{T_\pm}$
 for taking test function $H_\pm$ part. In our model test functions might appear alone, or multiplied with system factors or inside the integral. We define each case appearing in our model under this notation.

  In \Ep{53z1} the $\ome$ dependent test functions are $\bra f |\ome\ket$, $e^{-i\ome t}$ and $\bra \ome |g\ket$.
  It is related to $z$ pole in the lower half plane, so $H_-$ part of test functions should be taken. For the properties of $\big( \big)^{T_-}$, first we demand that it is equal to the $[\,\, ]^-$ operator when there are only test functions.  For example,
  \beqa
   \bigg( \bra f|\ome \ket \bigg)^{T_-} \equiv [\bra f|\ome \ket]^-, \,\,\,
   \bigg( \bra f|\ome \ket e^{-i\ome t} \bra \ome |g\ket \bigg)^{T_-} \equiv [\bra f|\ome \ket e^{-i\ome t} \bra \ome |g\ket]^-. \EQN{54fTm}
   \eeqa

  Next
  we want $\big( \big)^{T_-}$ to be linear to the test function parts. When the test functions are multiplied with non test function parts,we want
   \beqa
    \big( a_\ome \bra f|\ome\ket + b_\ome  \bra \ome |g\ket \big)^{T_-} = a_\ome \big( \bra f|\ome \ket \big)^{T_-} +
    b_\ome \big( \bra \ome |g\ket \big)^{T_-} \EQN{55lin}
    \eeqa
    where $a$ and $b$ are functions which belong to system part. Up to this point, $()^{T_+}$ can be defined similar way and we have the property
     \beqa
      & &\big(F \big)^{T_-} + \big( F \big)^{T_+} = F, \EQN{56Tcomp} \\
      & & \bigg( \big ( F \big)^{T_-} \bigg )^{T_+} =0 ,\,\,\, \bigg( \big ( F \big )^{T_-} \bigg )^{T_-} = \big ( F \big )^{T_-} \EQN{57orth}\\
      & & F = a_\ome \bra f|\ome\ket + b_\ome  \bra \ome |g\ket.
     \eeqa
 If \Ep{56Tcomp} \Ep{57orth} holds for other forms of $F$, $\big( \big)^{T_\pm}$ becomes complete and orthogonal like
 $[\,]^{\pm}$ operators. We try to keep this properties for all possible cases of $F$ in \Ep{53z1}. 

 Having the requirements of linearity, completeness and orthogonality in mind, let us examine \Ep{53z1} more closely. Since $z$ encircling contour can be arbitrarily close to $z$, we can replace
 \beqa
  A_\ome \rightarrow \frac{-N}{2\pi i (\ome-z)}  ,\,\,\, B_{\ome' \ome} \rightarrow B_{\ome' z},\,\,\, B^c_{\ome' \ome} \rightarrow B^c_{\ome' z}.  \EQN{56repl}
 \eeqa
 Note that we did not replace $\ome$ in the test functions.

   In \Ep{53z1} some terms have test functions inside the integration. With \Ep{44fH-}and \Ep{56repl} we rewrite \Ep{53z1} as
    \beqa
    & &\int_{C_{z}}d\ome \frac{-N}{2\pi i (\ome-z)} \bigg( \bra f|1\ket + \int_0^{\infty}d\ome' \frac{\bra f|\ome'\ket B_{\ome'z}}{\ome-\ome'+i\eps}\bigg) e^{-i\ome t}\bigg( \bra 1|g\ket + \int_0^{\infty}d\ome' \frac{\bra \ome'| g\ket B_{\ome'z}^{c}}{\ome-\ome'+i\eps}\bigg) \nonumber \\
    & &=\int_{C_{z}}d\ome \frac{-N}{2\pi i (\ome-z)} \bigg( \bra f|1\ket - 2\pi i [ \Theta (\ome) \bra f|\ome\ket B_{\ome z} ]^+
     \bigg) e^{-i\ome t}\bigg( \bra 1|g\ket -2 \pi i [ \Theta (\ome)  \bra \ome | g\ket B_{\ome z}^{c}   ]^+
     \bigg) \nonumber \\
      \EQN{57B}
     \eeqa
     where we define
\beqa
   \Theta (\ome) =\left\{ \begin{array}{ll}
     1\,\, & \mbox{for $\ome>0$} \\
     \frac{1}{2} \,\, &\mbox{for $\ome=0$} \\
     0 \,\, & \mbox{for $\ome<0$.}
     \end{array} \right.
   \eeqa
 Also, when there is no confusion and $\ome$ is the only variable we abbreviate
       \beqa
       \Theta(\ome) \rightarrow  \Theta,\,\,\, B_{\ome z} \rightarrow B_{z},\,\,\, B^c_{\ome z} \rightarrow B^c_{z},\,\,\,   \bra f|\ome\ket \rightarrow f,\,\,\, e^{-i\ome t} \rightarrow e,\,\,\, \bra \ome |g\ket \rightarrow g.
        \eeqa

       In \Ep{57B} we need to consider four cases, $e$, $e [ \Theta   B_{z} \, f ]^+$, $e[ \Theta   B^{c}_{z} \,g ]^+$ and $[\Theta B_{z} f ]^+\, e \, [\Theta B_{z}^{c} g ]^+$. We construct $\big( \big)^{T_\pm}$ for each case. For details, see appendix~\ref{App2}.

 We have
        \beqa
         & &\big(e \big)^{T_-} \equiv [e]^- = e^-, \EQN{63e}
       \eeqa
  \beqa
 & &\bigg(  e [\Theta B_{z} f ]^+ \bigg)^{T_-} \equiv  \Theta B_{z} [ e^- f]^- - e^- [\Theta B_{z} f ]^-, \EQN{75ef+} \\
 & &\bigg(  e [\Theta B^c_{z} g ]^+ \bigg)^{T_-} \equiv  \Theta B^c_{z} [ e^- g]^- - e^- [\Theta B^c_{z} g ]^- \EQN{76eg+}
  \eeqa   
  \beqa
   & &\bigg( [\Theta B_{z} f ]^+\, e \, [\Theta B_{z}^{c} g ]^+ \bigg)^{T_-} \nonumber \\
      & &\equiv \Theta B_{z} B^{c}_{z} \, [f e^- g]^-  - \Theta B_{z}B^{c}_{z}  [ [f e^-]^+ g]^- - \Theta B_{z} [f e^-]^- [\Theta B_{z}^{c} g ]^- \nonumber \\
      & &- \Theta B_{z} B^{c}_{z}[ f [e^- g]^+]^- - \Theta B_{z}^{c}[\Theta B_{z} f ]^-  [e^-g]^- + [\Theta B_{z} f ]^- e^- [\Theta B_{z}^{c} g ]^-. \EQN{66fgTm}
   \eeqa
  and
  \beqa
         & &\big(e \big)^{T_+}  = e^+, 
       \eeqa
       \beqa
 & &   \bigg( e [\Theta B_{z} f ]^+ \bigg)^{T_+} =e^+ [B_z f]^+ + B_z [e^- f]^+ \\
 & &\bigg( e [\Theta B^c_{z} g ]^+ \bigg)^{T_+} =e^+ [B^c_z g]^+ + B^c_z [e^- g]^+
 \eeqa
\beqa
     & &\bigg( [\Theta B_{z} f ]^+\, e \, [\Theta B_{z}^{c} g ]^+ \bigg)^{T_+}  \nonumber \\
        & &= [\Theta B_{z} f ]^+\, e^+ \, [\Theta B_{z}^{c} g ]^+  + \Theta B_{z} B_{z}^{c} [f e^- g ]^+
        -\Theta B_{z} B_{z}^{c} \big[ [f e^-]^+  g \big]^+  \nonumber \\
        & &+ \Theta B_{z}  [f e^-]^+ [\Theta B_{z}^{c} g ]^+  - \Theta B_{z} B_{z}^{c} \big[ [e^- g]^+  f \big]^+
        + \Theta B_{z}^{c} [ e^- g]^+ [\Theta B_{z} f ]^+. 
 \eeqa

 From \Ep{63e} to \Ep{66fgTm} we can also see the condition for $z$ complex eigenstates being remained as $z$ complex eigenstates when the above test function procedure is applied. Whenever the conditions
 \beqa
  e = e^-,\,\,\, [e^- f]^- = e^- f,\,\,\, [e^- g]^- = e^- g,\,\,\, [fe^- g]^- = fe^- g \EQN{71zcon}
 \eeqa
are all satisfied the $z$ complex eigenstate is remained as $z$ complex eigenstate. For some parameter (possibly $t$ or $x$) region where \Ep{71zcon} are not satisfied, $e^{-i z t}$ form might not be maintained and test function restricted form is not a $z$ complex eigenstate anymore. 

     For the third term of \Ep{55fg} we can proceed similarly. In that case the pole $z^{c.c}$ is in the upper half plane and $H_+$ parts of test functions should be taken. The relation
     \beqa
       & &\bra f|\phi_{z^{c.c}}\ket e^{-i z^{c.c} t} \bra \phit_{z^{c.c}}|g\ket  \nonumber \\
 & & = \int_{C_{z^{c.c}}}d\ome A_\ome^{c} \bigg(\bra f |1\ket + \int_0^{\infty}d\ome' \frac{\bra f|\ome'\ket B_{\ome'\ome}}{\ome-\ome'-i\eps}\bigg) e^{-i\ome t}\bigg( \bra 1|g\ket + \int_0^{\infty}d\ome' \frac{\bra \ome'|g\ket  B_{\ome'\ome}^{c}}{\ome-\ome'-i\eps}\bigg) \EQN{77z1cc}
     \eeqa
     again contains diverging terms and we remove these effects one by one. Like $z$ case we get
      \beqa
     & & \bigg(  e [ \Theta B_{z^{c.c}} f ]^- \bigg)^{T_+} \equiv \Theta B_{z^{c.c}} [ e^+ f]^+ - e^+ [\Theta B_{z^{c.c}} f  ]^+,  \\
     & &\bigg(  e [ \Theta B^c_{z^{c.c}} g ]^- \bigg)^{T_+} \equiv \Theta B^c_{z^{c.c}} [ e^+ g]^+ - e^+ [\Theta B^c_{z^{c.c}} g  ]^+, \EQN{74egTp} \\
     & &\bigg( [\Theta B_{z^{c.c}} f ]^-\, e \, [\Theta B_{z^{c.c}}^{c} g ]^- \bigg)^{T_+} \nonumber \\
     & &\equiv \Theta B_{z^{c.c}} B_{z^{c.c}}^c  \, [f e^+ g]^+  - \Theta B_{z^{c.c}} B_{z^{c.c}}^c  [ [f e^+]^- g]^+ - \Theta B_{z^{c.c}} [f e^+]^+ [\Theta B_{z^{c.c}}^{c} g ]^+ \nonumber \\
      & &- \Theta B_{z^{c.c}} B_{z^{c.c}}^c [ f [e^+ g]^-]^+ -\Theta  B_{z^{c.c}}^c [B_{z^{c.c}} f ]^+  [e^+g]^+ + [\Theta B_{z^{c.c}} f ]^+ e^+ [\Theta B_{z^{c.c}}^{c} g ]^+. \EQN{83z1cc}
      \eeqa
    and
    \beqa
      & & \bigg(  e [ \Theta B_{z^{c.c}} f ]^- \bigg)^{T_-} \equiv e^-[\Theta B_{z^{c.c}} f]^- + \Theta B_{z^{c.c}} [ e^+ f]^- , \\
      & &\bigg(  e [ \Theta B^c_{z^{c.c}} g ]^- \bigg)^{T_-} \equiv e^-[\Theta B^c_{z^{c.c}} g]^- + \Theta B^c_{z^{c.c}} [ e^+ g]^-, \\
       & &\bigg( [\Theta B_{z^{c.c}} f ]^-\, e \, [\Theta B_{z^{c.c}}^{c} g ]^- \bigg)^{T_-} \nonumber \\
     & &\equiv  [\Theta B_{z^{c.c}} f ]^-\, e^- \, [\Theta B_{z^{c.c}}^{c} g]^-    
     + \Theta B_{z^{c.c}} B^c_{z^{c.c}} [f e^+ g]^- - \Theta B_{z^{c.c}} B^c_{z^{c.c}} \big [ [ f e^+]^- g \big ]^- \\
     & &+ \Theta B_{z^{c.c}} [f e^+]^- [\Theta B^c_{z^{c.c}} g]^- - \Theta B_{z^{c.c}}  B^c_{z^{c.c}} \big[ [e^+ g]^-
     f\big]^- + \Theta B^c_{z^{c.c}} [e^+ g]^- [ \Theta B_{z^{c.c}} f]^-.
    \eeqa
Next section we apply above formulations to the well known specific model for actual calculations of complex eigenstates.

 \section{A specific example: Friedrichs model  }
 In this section we introduce a decay model known as Friedrichs model as a specific example of our more general formulation.
 We review its eignestates, poles and complex eigenstates in this section and their specific forms.
  This model has a same form as \Ep{3H0V} and describes a discrete system interacting with a 1D continuous scalar field. The interaction potential has a simple form and exact analytic eigenstates as well as complex eigenstates are known. Since this model has the same from as \Ep{3H0V} and serves as a specific example, the same symbols $H$, $V$ etc. will be used when there is no confusion.

  The Friedrichs Hamiltonian is given by
 \beqa
\Hb = \ome_1 |1\ket\bra 1| + \int_{-\infty}^{\infty} dk \; \ome_k
|k\ket \bra k| + \lam
 \int_{-\infty}^{\infty} dk \; \vb_k (|1\ket \bra k| + |k\ket \bra 1|). \EQN{19F}
 \eeqa
 The state $|1\ket$ represents the discrete state and the state
$|k\ket$ represents a continuous scalar field of momentum $k$. In spatial representation
\beqa
\bra x|k\ket= \frac{1}{\sqrt{2 \pi}} e^{ikx} \EQN{78xk}
\eeqa
which is delta normalized. For simplicity we set $c=\hbar=1$.

The interaction Hamiltonian is
\beqa
\Vb = \lam
 \int_{-\infty}^{\infty} dk \; \vb_k (|1\ket \bra k| + |k\ket \bra 1|).
\eeqa
The interaction term $\vb_k (|1\ket \bra k|
+ |k\ket \bra 1|)$ stands for the transition from $1$ state to $k$
state and from $k$ state to $1$ state. Coupling constant $\lam$ is a parameter indicating interaction strength.

The energy of the lowest field state is chosen to be zero; $\ome_1$  is
the eigenenergy of the discrete state and field $\ome_k$ dispersion relation is
\beqa
\ome_k = |k|.
\eeqa
 The dimensionless constant $\lam$ is chosen to be small
($\lam \ll 1$) such that the decay phenomena is well noticable. We shall consider a specific form of the
interaction potential
\begin{eqnarray}
  \vb_k  =  \frac{\sqrt{\ome_k}}{ 1 +(\ome_k/M)^2}.
 \EQN{4}
\end{eqnarray}
The constant $M^{-1}$ determines the range of the interaction and
gives an ultraviolet cutoff. Since we dealing with analytic continuations of energy,  we choose the branch cut of $\sqrt{\ome}$ in \Ep{4} as negative real line so that analytic continuation from positive real line does not pose any problems.

  The Hamiltonian $\Hb$ in \Ep{19F} can be more simplified. From the dispersion relation $\ome_k = |k|$, the free-Hamiltonian
 eigenstates $|k\ket$ and $|-k\ket$ have the same eigenvalue $\ome_k$. We
 remove this degeneracy by rewriting the Hamiltonian
  \beqa
\Hb = \ome_1 |1\ket \bra 1| + \int_{0}^{\infty} dk\; \ome_k
   \left(|S_{k}\ket \bra S_{k}| + |A_{k}\ket \bra A_{k}| \right)
    + \int_{0}^{\infty} dk\;\sqrt{2}\lam \vb_{k}
   (|1\ket \bra S_{k}|+|S_{k}\ket\bra 1|)
   \EQN{5-1}
 \eeqa
 where
\beqa
  |S_{k}\ket  \equiv \frac{1}{\sqrt{2}} (|k\ket + |-k\ket),\qquad
  |A_{k}\ket  \equiv \frac{1}{\sqrt{2}} (|k\ket - |-k\ket). \EQN{5-2}
  \eeqa
 From \Ep{5-1} we see that the discrete eigenstate $|1\ket$ only
 interacts with the symmetric field eigenstate $|S_{k}\ket$.
  From now on, we concentrate only on the
 discrete atom state and the symmetric field states which are related to the decay.
 \beqa
  |\ome\ket \equiv |S_{k}\ket,\;\;\; v_{\ome} \equiv
  \sqrt{2}\, \vb_{\ome},\,\,\,  \mbox{$\ome \ge 0$}    \EQN{2-6}
  \eeqa
 and  ignoring $|A_{k}\ket$ related
 terms we get the Hamiltonian
  \beqa
 & &H \equiv  H_{0}+ V \nonumber \\
 & &=\ome_1 |1\ket\bra 1| + \int_0^{\infty}d\ome\, \ome |\ome\ket \bra \ome| + \int_0^{\infty}d\ome\, \lam v_{\ome}
(|1\ket\bra \ome|+ |\ome\ket\bra 1|), \EQN{Fmodel} \\
   & &H_{0}\equiv \ome_1 |1\ket \bra 1| + \int_{0}^{\infty} d\ome\; \ome
   |\ome\ket \bra \ome|,\;\;  V\equiv \int_{0}^{\infty} d\ome\; v_{\ome}
   (|1\ket \bra {\ome}|+|{\ome}\ket\bra 1|).
   \eeqa

  This Hamiltonian has an exact diagonalized form and various properties of the exact solution have been
   analyzed \cite{PPT}. When $\ome_1 > \int_{0}^{\infty} d\ome
 \lam^2 v_\ome^2 /\ome$,
  We can write
  \beqa
  H =  \int_0^{\infty}d\ome \ome |F^+_{\ome}\ket \bra F^+_{\ome}| =
   \int_0^{\infty}d\ome \ome |F^-_{\ome}\ket \bra F^-_{\ome}| \EQN{(2)}
  \eeqa
 where
  \beqa
  & &|F^+_\ome\ket= |\ome\ket + \frac{\lam v_\ome}{\eta^+(\ome)} |1\ket  + \frac{\lam v_\ome}{\eta^+(\ome)}\int_0^{\infty}
  d\ome' \frac{\lam v_{\ome'} |\ome'\ket}{\ome-\ome'+i\eps} \EQN{31Fpome} \\
  & &|F^-_\ome\ket=|\ome\ket + \frac{\lam v_\ome}{\eta^-(\ome)} |1\ket  + \frac{\lam v_\ome}{\eta^-(\ome)}\int_0^{\infty}
  d\ome' \frac{\lam v_{\ome'} |\ome'\ket}{\ome-\ome'-i\eps} \EQN{32Fmome}
  \eeqa
 with
  \beqa
 \eta^{\pm} (\ome) \equiv \ome-\ome_1 -\int_0^{\infty}d\ome\frac{\lam^2
 v_{\ome}^2}{\ome^{\pm}-\ome}= \ome-\ome_1 - P\int_0^{\infty}d\ome\frac{\lam^2
 v_{\ome}^2}{\ome-\ome} \pm i\pi \lam^2 v_{\ome}^2    \EQN{94eta}
 \eeqa
 where $P\int$ means Cauchy principal integral.

 In \Ep{94eta},
 $1/(\ome^{\pm}-\ome)$ means that $z$ is analytically
 continued from above ($+$) or below ($-$). For real $\ome$, it can be
 understood as $\ome^{\pm} \equiv \ome\pm i\eps$, where $\eps >0$ is
 infinitesimal.  We can choose $+$ branch or $-$ branch for the diagonalized solution and
 these two sets of eigenstates independently satisfy the eigenvalue equation as well as
  the orthonormality and completeness relations
  \beqa
 H|F_\ome^{\pm}\ket = \ome |F_\ome^{\pm}\ket, \;\;\;
   \bra F_\ome^\pm |F_{\ome'}^\pm\ket = \delta (\ome-\ome'),\;\;\; \int_{0}^{\infty}d\ome\;
   |F_\ome^\pm\ket \bra F_\ome^\pm| = 1
   \eeqa
  It is noted in \EQN{(2)} no discrete eigenstate is present. The discrete state $|1\ket$ is represented
   as sum of continuum state and it decays into continuum if there is no initial field.

 Actually these sets of solutions are the solutions of Lippmann-Schwinger equation we discussed in section~\ref{sec:Lippmann}.
From  \Ep{31Fpome} we have
 \beqa
  \bra 1 |F_\ome^+\ket = \frac{\lam v_\ome}{\eta^+(\ome)}, \,\,\,\bra 1 |F_\ome^-\ket = \frac{\lam v_\ome}{\eta^-(\ome)},\,\,\,
  \bra \ome'|V|F_\ome^+\ket = \lam v_{\ome'} \frac{\lam v_\ome}{\eta^+(\ome)}
 \eeqa
  and direct substitution into \Ep{5F+} shows that $|F_\ome^+\ket$ is indeed the solution.

  The S matrix of Friedrichs model  is given by
  \beqa
   & &S(\ome) = \frac{\eta^-(\ome)}{\eta^+(\ome)} = 1 -2 \pi i \frac{\lam^2 v_\ome^2}{\eta^+(\ome)}, \nonumber \\
   & &  |F_\ome^+\ket= S(\ome) | F_\ome^-\ket.
  \eeqa
   In Friedrichs model setting \Ep{20Sep} is written as
   \beqa
   & &|F_\ome^+\ket\bra F_\ome^+| \nonumber \\
   & &= |\ome\ket\bra \ome| \nonumber \\
   & &+ \frac{1}{2\pi i}\frac{1}{\eta^-(\ome)} \bigg( |1\ket+
   \int_0^{\infty}d\ome' \frac{\lam v_{\ome'}|\ome'\ket}{\ome-\ome' - i\eps}\bigg) \bigg( \bra 1 |+
   \int_0^{\infty}d\ome' \frac{\lam v_{\ome'}\bra \ome'|}{\ome-\ome' - i\eps}\bigg) \nonumber \\
   & &+ \frac{-1}{2\pi i}\frac{1}{\eta^+(\ome)} \bigg( |1\ket+
   \int_0^{\infty}d\ome' \frac{\lam v_{\ome'}|\ome'\ket}{\ome-\ome' + i\eps}\bigg) \bigg( \bra 1|+
   \int_0^{\infty}d\ome' \frac{\lam v_{\ome'}\bra \ome' |}{\ome-\ome' + i\eps}\bigg). \EQN{94Fr}
   \eeqa
    Each pole contribution is clearly separated in \Ep{94Fr}. Let us write the pole of
     $1/\eta^+(\ome)$ closest to $\ome_1$ as $z$ and
   and that of $1/\eta^-(\ome)$ closest to $\ome_1$ as  $z^{c.c}$ , in lower and upper half planes, respectively.
  The complex eigenstates can be obtained by analytically continuing real energy spectrum into complex plane, using those poles.

The explicit forms of complex
   eigenstates are written as
   \beqa
   & & |\phi_{z}\ket = N^{1/2} \bigg( |1\ket + \int_0^{\infty}d\ome \frac{ \lam v_\ome |\ome\ket}{ z^+ -\ome}\bigg),
    \EQN{(6)} \\
   & & \bra \phit_{z}| = N^{1/2} \bigg( \bra 1|
   + \int_0^{\infty}d\ome \frac{ \lam v_\ome \bra\ome|}{ z^+ -\ome}\bigg).
   \eeqa
 This complex eigenvectors have eigenvalues
  \beqa
  & &H|\phi_{z}\ket = z |\phi_{z}\ket,\;\;\;  \\
  & &\bra \phit_{z}| H = \bra \phit_{z}| z,\;\;\;
  \eeqa

 Similarly we have $z^{c.c}$ complex eigenstates
 \beqa
 & & |\phi_{z^{c.c}}\ket = (N^{c.c})^{1/2} \bigg( |1\ket + \int_0^{\infty}d\ome \frac{ \lam v_\ome |\ome\ket}{ {z^{c.c}}^- -\ome}\bigg),
     \\
   & & \bra \phit_{z^{c.c}}| = (N^{c.c})^{1/2} \bigg( \bra 1|
   + \int_0^{\infty}d\ome \frac{ \lam v_\ome \bra\ome|}{ {z^{c.c}}^- -\ome}\bigg).
  \eeqa
  and
   \beqa
    & &H|\phi_{z^{c.c}}\ket = z^{c.c} |\phi_{z^{c.c}}\ket,\;\;\;  \\
  & &\bra \phit_{z^{c.c}}| H = \bra \phit_{z^{c.c}}| z^{c.c},\;\;\;
   \eeqa

    Without test function considerations, these complex eigenstates show exponential blowup behavior.
    If time evolution operator $e^{-iHt}$ is
   applied to complex eigenvector $|z\ket$, it gives $e^{-iz t}$ factor which grows exponentially for negative time.
   Exponential growth also appears in spatial domain. If $\bra x|z\ket$ is considered, it yields
    \beqa
    \bra x|\phi_{z}\ket = N \int_0^{\infty}d\ome \frac{ \lam v_\ome \bra x |\ome\ket}{ z^+ -\ome}
    \eeqa
   and its spatial feature is shown in figure 1. Exponential growth $e^{ iz |x|}$ is shown as $|x|$ increases.

Next section we compare numerical total Hamiltonian time evolution with complex eigenstates.
It is shown that in certain ranges of spacetime
 the complex eignestate component is very close to the actual field. Test function restriction of complex eigenstates clearly shows those physically meaningful regions
 of spacetime.

 \section{ Comparison of test function restricted complex eigenstate with total Hamiltonian evolution}

 In this section we do the numerical simulation of total Hamiltonian time evolution, and compare results with test function restricted complex eigenstate. Our original goal is to find physically meaningful regions of complex eigenstate, and comparison with total time evolution should justify our construction.

  Numerical setup is as follows. We choose $\ome_1 =2$, $\lam = 0.1$. Continuous field modes are discretized inside a box with size $L=100$. Total number of discretized field mode is $N=1200$. The energy cutoff constant in \Ep{4}
  is chosen as $M=5$.
  With this setup discretized $\ome$ becomes
  \beqa
   \ome_n = \frac{2 \pi}{L}n.
   \eeqa
   With the box normalization of size $L$, delta functions correspond to Kronecker delta as
   \beqa
   \delta (\ome-\ome')  \longleftrightarrow \frac{L}{2\pi}\delta_{\ome,\ome'}
   \eeqa
   If we require orthonormal relations for energy eigenstates in discrete case, we get
   \beqa
    & &\bra \ome_n|\ome_{n'}\ket = \delta_{n,n'}.
    \eeqa
    From the correspondences
    \beqa
   \sum_k  \bra \ome_n|x_k\ket\bra x_k|\ome_{n'}\ket \Delta x \leftrightarrow \int dx \bra \ome|x\ket\bra x|\ome\ket,\,\,\,
   \sum_n V_n^2   \leftrightarrow \int d\ome v_\ome^2
    \eeqa
    we have
    \beqa
    & & \bra \ome_n |x\ket = \sqrt{ \frac{2}{L} } \cos (\ome_n x), \\
    & &  V_n = 2 \sqrt{ \frac{\pi}{L}} \big ( \frac{ \ome_n^{1/2}}{ (\ome_n /5)^2 +1} \big).
   \eeqa
 Total discrete Hamiltonian becomes
\beqa
  H = \ome_1 |1\ket\bra 1| + \sum_n \ome_n |\ome_n\ket\bra \ome_n|
   + \lam \sum_n V_n (|\ome_n\ket\bra 1| + |1\ket\bra \ome_n|).
   \eeqa

 We directly calculate the time evolution operator $e^{-iH t}$ using fourth order Crank-Nicolson method \cite{CN1,CN2}.
 Numerical accuracy is checked in multiple ways since this exactly diagonalized form of discrete Hamiltonian and analytic solution for continuous case are both known.

  Let us compare numerical results with test function restricted complex eigenstates in  \Ep{75ef+} - \Ep{66fgTm}. 

\begin{figure}[hbp]
\includegraphics[width=8 cm,height= 4 cm]{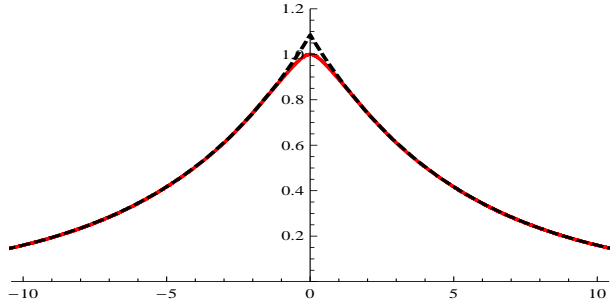}
\caption{ $|\bra 1|e^{-iHt}|1\ket|^2 $ (solid) versus
$ \bigg| \big( \bra 1|e^{-iHt}|1\ket \big)^T \bigg|^2$ (dashed) plots.
They show very close agreements. When test functions are restricted both pole components give dominant
contributions in their physically meaningful regions. }
\label{11tplots}
\end{figure}

   First we consider the case $\bra f| =\bra 1|$, $|g\ket = |1\ket$.   From \Ep{29comp} we have
   \beqa
    \bra 1|e^{-iHt}|1\ket =  \bra 1  |\phi_{z^{c.c}}\ket e^{-i z^{c.c} t} \bra \phit_{z^{c.c}}|1\ket +
 \bra 1 |\phi_{z}\ket e^{-i z t} \bra \phit_{z}|1\ket+ (\mbox{rest}). \EQN{118fg11}
    \eeqa
  In \Ep{118fg11} the test functions are not restricted, so the term containing $e^{-i z^{c.c} t}$ grows for positive $t$ and the term containing $e^{-i z t}$ grows for negative $t$, exponentially. To avoid this we take test function restricted form
  \beqa
   & & \bigg( \bra 1|e^{-iHt}|1\ket \bigg)^T  \nonumber \\
   & &=\int_{C_{z}} d\ome  \bigg(  \frac{-1}{2 \pi i}\frac{e^{-i \ome t}}{ \eta^+(\ome) } \bigg)^{T_-} +
 \int_{C_{z^{c.c}}} d\ome\bigg( \frac{1}{2 \pi i}\frac{e^{-i \ome t} }{ \eta^-(\ome) } \bigg)^{T_+} \nonumber \\
 & &= \Theta(t) N e^{-i z t}+ \Theta(1-t) N^{c.c} e^{-i z^{c.c} t}. \EQN{119fg11T}
   \eeqa
   \Ep{119fg11T} shows clear distinction between $z$ pole component and $z^{c.c}$ component. For $t<0$ only
    $z^{c.c}$ pole component contributes and for $t>0$ only $z$ pole component contributes. Figure~\ref{11tplots} shows the comparison between $| \bra 1 |e^{-i\ome t}|1\ket |^2$ and
    $ \bigg |\bigg( \bra 1|e^{-iHt}|1\ket \bigg)^T \bigg|^2 $.
 Test function restricted complex eigenstates (thick dashed line) show very good agreement with $|\bra 1|e^{-iHt}|1\ket|^2 $  time evolution (solid line) for their respective regions.

\begin{figure}[hbp]
\includegraphics[width=8 cm,height= 4 cm]{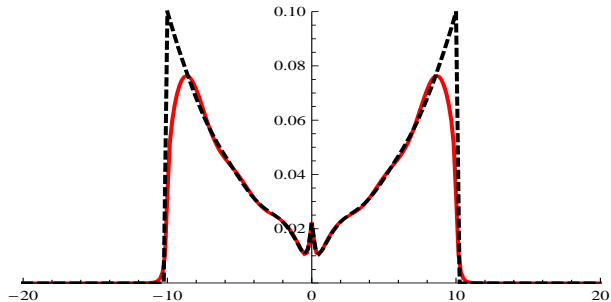}
\caption{ $|\bra x|e^{-iHt}|1\ket|^2 $ (solid) versus
$ \bigg | \big( \bra 1 | e^{-i H_F t} |x\ket   \big)^{T} \bigg |^2$ (dashed) plots for $t=10$.
They show close agreements for physically meaningful regions. Characteristic emitting decay fields are shown in both
plots and outside the causal region remaining fields are very small.  }
\label{x1tplots}
\end{figure}

Next we consider the case of $\bra f| = \bra 1|$ and $|g\ket$ is field component. This is related to the
   emission and absorption with discrete state.
   To see the effect of field appearing in space we choose $|g\ket = |x\ket$. From \Ep{78xk} $\bra \ome|x\ket$ is given by
 \beqa
 \bra \ome|x\ket=\frac{1}{\sqrt{\pi}} \cos(\ome x)
 \eeqa
and we have for $t>0$
 \beqa
 & &\bigg( \bra 1 | e^{-i H t} |x\ket   \bigg)^{T} \nonumber \\
& &= \int_{C_{z}} d\ome \bigg(  \frac{-1}{2\pi i}\frac{1}{\eta^+(\ome)} e^{-i\ome t}
   \int_0^{\infty}d\ome' \frac{\lam v_{\ome'}\bra \ome' |x\ket }{\ome-\ome' + i\eps} \bigg)^{T_-} \nonumber \\
  & &+ \int_{C_{z^{c.c}}} d\ome \bigg( \frac{1}{2\pi i}\frac{1}{\eta^-(\ome)} e^{-i\ome t}
   \int_0^{\infty}d\ome' \frac{\lam v_{\ome'}\bra \ome'|x\ket }{\ome-\ome' - i\eps} \bigg)^{T_+} \nonumber \\
 \nonumber \\
  & &=  \int_{C_{z}} d\ome \bigg(  \frac{-1}{2\pi i}\frac{1}{\eta^+(\ome)} e^{-i\ome t}
   \int_0^{\infty}d\ome' \frac{\lam v_{\ome'}\bra \ome' |x\ket }{\ome-\ome' + i\eps} \bigg)^{T_-} \nonumber \\
  & &=   \sqrt{\pi} i N  \Theta (t) \bigg( v_{z} (\Theta(t-|x|) e^{-i z (t-|x|)} + e^{-i z (t+|x|)})
  - e^{-i z t} [ \Theta(\ome) v_\ome \bra \ome |x\ket ]^- \bigg) \EQN{114f1gx}
  \eeqa
  In \Ep{114f1gx} $z^{c.c}$ contribution is zero for $t>0$ as we seen in \Ep{74egTp}. Only $e^+$ parts are in \Ep{74egTp} and $[e^{-i\ome t} ]^+$ is zero for $t>0$.
The result in \Ep{114f1gx} is compared with $\bra 1 |e^{-i H t}|x\ket $ in absolute square values in Figure \ref{x1tplots}. $ \bigg | \big( \bra 1 | e^{-i H_F t} |x\ket   \big)^{T} \bigg |^2$ shows close agreement with total
time evolution. Emitting fields are clearly shown and match well in physically meaningful region, and outside the
causal region both fields are very small. Test function restricted complex eigenstates capture physical
features well.

  $\bra f| =\bra x|$ and $|g\ket = |1\ket$ case is the almost same as \Ep{114f1gx} with only bras and kets are exchanged. When absolute values are taken the results are same as \Ep{114f1gx} so we do not elaborate here.

 Finally we consider the case of  both $\bra f|$ and $|g\ket$ are field components. From the decomposition of \Ep{94Fr} we see that $\bra f|\ome\ket e^{-i\ome t} \bra \ome |g\ket$ component does not vanish. To see the free field effect and scattering effect, let us examine the case of $\bra f| = \bra x_1 |$
    and $|g\ket = |x_2\ket$. The correlations between fields are shown when we examine
    $\bra x_1 |e^{-iH_F t} |x_2\ket$.

    In \Ep{94Fr} we divided the complete set $|F_\ome^+\ket\bra F_\ome^+|$ as free field contribution, lower half plane pole contribution and upper half plane pole contribution. The free field contribution is
     \beqa
      & &\int_0^{\infty} d\ome \bra x_1|\ome\ket e^{-i\ome t} \bra \ome|x_2\ket
      = \frac{1}{\pi}\int_0^{\infty} d\ome \cos (\ome x_1) e^{-i\ome t} \cos (\ome x_2) \nonumber \\
      & &= \frac{1}{4} ( \delta( x_1 -x_2-t) + \delta(x_1+x_2 -t) + \delta(x_1-x_2 +t) +   \delta(x_1+x_2 +t) )
      \nonumber \\
      & &+\frac{i}{\pi} \int_0^{\infty} d\ome \cos (\ome x_1) \sin (\ome t) \cos (\ome x_2).  \EQN{122}
      \eeqa
     It yields four delta function and an integration involving $\sin (\ome t)$. In spacetime picture the delta
     functions are highly localized, and the integral involving $\sin (\ome t)$ becomes very small
as $t$ becomes large. So the contributions from free field correlations are very distinctive compared to other
contributions.

 Next we examine the contributions from poles. There are two poles, one in the lower half plane and the other in the upper half plane. As we can see in \Ep{66fgTm} and \Ep{83z1cc} all $(\,\,)^{T_-}$ parts contains $[e^{-i\ome t}]^-$
 and all $(\,\,)^{T_+}$ contains $[e^{-i\ome t}]^+$. This means for $t>0$ only  $(\,\,)^{T_-}$ parts contribute and
   $z^{c.c}$ parts are only for $t<0$.

 So for $t>0$ we calculate $z$ encircling integration with $( \,\,)^{T_-}$
 and compare them with the total time evolution $\bra x_1 |e^{-iH_F t} |x_2\ket$.

 From \Ep{66fgTm},
  \beqa
    & &\int_{C_{z}} d\ome  \big( \bra x_1 |F_\ome^+\ket e^{-i \ome t}\bra \ome|x_2\ket \big)^{T_-} \nonumber \\
   & &\int_{C_{z}} d\ome \bigg( \frac{-1}{2\pi i}\frac{1}{\eta^+(\ome)}
   \int_0^{\infty}d\ome' \frac{\lam v_{\ome'}\bra x_1 |\ome'\ket}{\ome-\ome' + i\eps}
  e^{-i\ome t} \int_0^{\infty}d\ome' \frac{\lam v_{\ome'}\bra \ome' |x_2\ket}{\ome-\ome' + i\eps} \bigg)^{T_-} \nonumber \\
  & & = \frac{N_1}{\pi} \lam^2 (2\pi i)^2  \bigg( [ \Theta(\ome) v_\ome \cos (\ome x_1) ]^+ e^{-i \ome t}
    [\Theta(\ome) v_\ome \cos (\ome x_2) ]^+ \bigg)^{T_-}_{z} \nonumber \\
   & & =\frac{N_1}{\pi} \lam^2 (2\pi i)^2 \Theta(t)  \bigg( v_{z}^2 [ \cos (\ome x_1) e^{-i \ome t} \cos (\ome x_2) ]^-
   - v_{z}^2 \big[ [  \cos (\ome x_1) e^{-i \ome t} ]^+  e^{-i \ome t} \big]^-  \nonumber \\
   & &- v_{z} [ \cos (\ome x_1)  e^{-i \ome t} ]^- [\Theta(\ome) v_\ome \cos (\ome x_2) ]^-
   - v_{z}^2 \big[  \cos (\ome x_1) [ e^{-i \ome t} \cos (\ome x_2)]^+ \big]^-  \nonumber \\
   & &-v_{z}  [ \Theta(\ome) v_\ome \cos (\ome x_1) ]^- [ e^{-i\ome t}  \cos (\ome x_2) ]^- \nonumber \\
   & &+ [ \Theta(\ome) v_\ome \cos (\ome x_1) ]^- e^{-i \ome t}
    [\Theta(\ome) v_\ome \cos (\ome x_2) ]^- \bigg)_{\ome=z} \nonumber \\
    & &= \frac{N_1}{\pi} \lam^2 (2\pi i)^2 \Theta(t) \bigg( \frac{v_{z}^2}{4} ( \Theta(t-|x_1 |-|x_2|) e^{-i z (t-|x_1| -|x_2| ) } + e^{-i z(t+|x_1|+|x_2|)} ) \nonumber \\
    & &- \frac{v_{z}}{4 \pi} (  \Theta (t-|x_1|) e^{-iz (t-|x_1|)} + e^{-iz (t+ |x_1|)} )
    \int_0^{\infty} d\ome' \frac{v_{\ome'} \cos (\ome x_2)}{z -\ome'} \nonumber \\
    & &- \frac{v_{z}}{4 \pi} \int_0^{\infty} d\ome' \frac{v_{\ome'} \cos (\ome x_1)}{z -\ome'} ( \Theta (t-|x_2|) e^{-iz (t-|x_2|)} + e^{-iz (t+ |x_2|)} \nonumber \\
    & &+ \frac{1}{4 \pi } \int_0^{\infty} d\ome' \frac{v_{\ome'} \cos (\ome x_1)}{z-\ome' } e^{-i z t} \int_0^{\infty} d\ome' \frac{v_{\ome'} \cos (\ome x_2)}{z -\ome'}
    \bigg). \EQN{123fgT}
  \eeqa
 Among various terms in RHS of \Ep{123fgT}, the first one is dominant. The physical meaning of this term is quite clear. Roughly, we can write
\beqa
  [ \Theta(\ome) v_\ome \cos (\ome x_1) ]^+ \approx v_\ome \frac{e^{i\ome |x_1|}}{2},\,\,\,
   [\Theta(\ome) v_\ome \cos (\ome x_2) ]^+ \approx v_\ome \frac{e^{i\ome |x_2|}}{2}
  \eeqa
 and from \Ep{78xk} we can interpret the outgoing wave from the point $x=0$ and the incoming wave toward $x=0$
 as
 \beqa
 \bra x_{out} |\ome\ket \propto e^{i\ome |x|},\,\,
 \bra x_{in} |\ome \ket  \propto e^{-i\ome |x|}.
 \eeqa
 Then in \Ep{123fgT}
 \beqa
 & &\bigg( [ \Theta(\ome) v_\ome \cos (\ome x_1) ]^+ e^{-i \ome t}   [\Theta(\ome) v_\ome \cos (\ome x_2) ]^+ \bigg)^{T_-}_{z} \nonumber \\
 & &\approx  C \int_0^{\infty}d\ome \bra {x_1} |\ome\ket_{out}\,\frac{e^{-i\ome t} }{\ome-z}  \, \bra x_2|\ome \ket_{in} \approx C' \Theta (t-|x_1|-|x_2|) e^{-iz (t-|x_1|-|x_2|)}
 \EQN{126inout}
  \eeqa
 where $C$ and $C'$ are constants. This shows that the dominant term can be interpreted as the correlation between incoming field at $x_2$ and outgoing field at $x_1$ through scattering at $x=0$. The time for the incoming field
 at $x_2$ scatters at $x=0$ point and changes to the outgoing field at $|x_1|$ is $t= |x_1|+|x_2|$. This correlation occurs at the resonance frequency $\ome=z$.

\begin{figure}[hbp]
\includegraphics[width=10 cm,height= 5 cm]{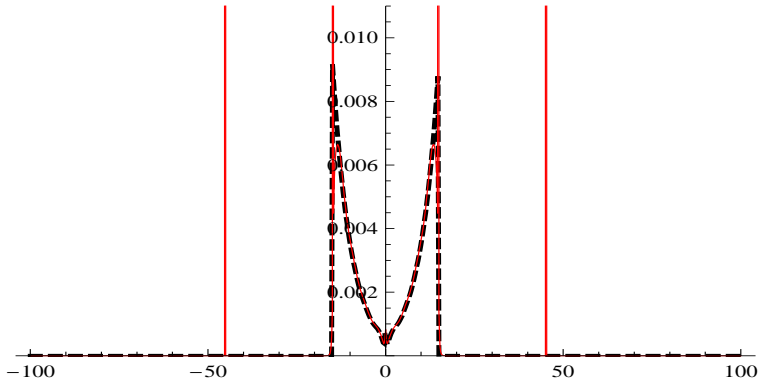}
\caption{ $|\bra x_1 |e^{-iHt}|x_2 \ket|^2 $ (solid) versus
$\bigg|\int_{C_{z}} d\ome  \big( \bra x_1 |F_\ome^+\ket e^{-i \ome t}\bra \ome|x_2\ket \big)^{T_-} \bigg |^2 $ (dashed) plots for $x_1$ with $t=30$ and $x_2= 15$. Four sharp peaks correspond to the numerical delta functions. Besides delta functions,
the scattering fields correlation shows close agreements.   }
\label{x1x2plots}
\end{figure}

  Figure~\ref{x1x2plots} shows the comparison between total time evolution $| \bra x_1 |e^{-iHt}|x_2 \ket |^2$ and

 $\bigg|\int_{C_{z}} d\ome  \big( \bra x_1 |F_\ome^+\ket e^{-i \ome t}\bra \ome|x_2\ket \big)^{T_-} \bigg |^2 $. Apart from delta functions which come from free fields component, the complex eigenstates with restricted test function captures the correlation between scattering fields well.

 The results show that choice of Hardy class
 functions as test functions of complex eigenstate selects only causal part of complex eigenstate and removes unphysical divergences.

 \section{Conclusion}
 We studied complex spectral representation in terms of the solution set of Lippmann-Schwinger equation. For a model of a discrete state and simple energy continuum it is shown that the complete solution set can be decomposed of a free field set, a set containing a lower half plane complex pole of Green's function and a set containing upper half plane complex pole of Green's function (\Ep{20Sep}). From this decomposition the complex eigenstates from both poles are constructed.
  To remove unphysical behaviors of complex eigenstates in spacetime, test functions are restricted. We consider
  Hardy class functions of real line for the restrictions of test functions. Decomposition formula of a square integrable function into Hardy class above and below functions is presented (\Ep{47Hd} and \Ep{44fH-}) and applied to upper half plane complex eignestate and lower half plane complex eignestate, respectively.

   To get physically meaningful results of time evolving complex eigenstates, the initial and final conditions as well as time evolution operators become subject to the test function restriction. Detailed decomposition formulas for test function restrictions are presented in section~\ref{sec:Htest}.
    When this results are applied to a specific model and test function restricted complex eigenstates are compared to the total time evolution, the complex eigenstate components show close resemblance to the total time evolution of decaying field. This demonstrates that test function restricted complex eigenstates capture the essence of decaying phenomena quite well.

   \begin{acknowledgments}
   Author thanks to Gonzalo Ordonez for helpful discussions and kind considerations.
This research was supported by the Academic Research fund of Hoseo University in 2013 (2013-0085).
\end{acknowledgments}

\appendix
\section{Derivation of \Ep{20Sep}} \label{App1}
 In this appendix we derive \Ep{20Sep}.
 First note that using \Ep{18SFp} and \Ep{23S1},
 \beqa
  & &\bra \Psi_\ome^+ |1\ket\bra 1|\Psi_\ome^+\ket =(\bra \Psi_\ome^+ |1\ket-\bra 1|\Psi_\ome^+\ket)  \frac{\bra \Psi_\ome^+ |1\ket\bra 1|\Psi_\ome^+\ket}{\bra \Psi_\ome^+ |1\ket-\bra 1|\Psi_\ome^+\ket} \nonumber \\
 & & = (\bra \Psi_\ome^+ |1\ket-\bra 1|\Psi_\ome^+\ket)  \frac{\bra 1|\Psi_\ome^+\ket}{1-\frac{\bra 1|\Psi_\ome^+\ket}{\bra \Psi_\ome^+|1\ket}  } = (\bra \Psi_\ome^+ |1\ket-\bra 1|\Psi_\ome^+\ket) \frac{ \bra 1|\Psi_\ome^+\ket   }
 {  2\pi i \bra \ome |V|\Psi_\ome^+\ket    } \nonumber \\
 & &=  (\bra \Psi_\ome^+ |1\ket-\bra 1|\Psi_\ome^+\ket)  \frac{\bra \Psi_\ome^+|1\ket}{\frac{\bra \Psi_\ome^+|1\ket}{\bra 1|\Psi_\ome^+\ket }-1  } = (\bra \Psi_\ome^+ |1\ket-\bra 1|\Psi_\ome^+\ket) \frac{ \bra \Psi_\ome^+|1\ket   }
 {  2\pi i \bra \Psi_\ome^+|V|\ome\ket    }. \EQN{a1d}
 \eeqa
  In \Ep{a1d} we see that
  \beqa
\frac{ \bra 1|\Psi_\ome^+\ket   }
 {  \bra \ome |V|\Psi_\ome^+\ket    } = \frac{ \bra \Psi_\ome^+|1\ket   }
 {  \bra \Psi_\ome^+|V|\ome\ket    }. \EQN{a2F}
  \eeqa
 From the expansion
 \beqa
  & &|\Psi_\ome^+\ket\bra \Psi_\ome^+| \nonumber \\
  & &= \bigg( |\ome\ket + |1\ket\bra 1|\Psi_\ome^+\ket + \int_0^{\infty}d\ome' \frac{|\ome'\ket\bra\ome'|V|\Psi_\ome^+\ket}{\ome-\ome'+ i \eps} \bigg)
   \bigg( \bra \ome | +\bra \Psi_\ome^+|1\ket\bra 1| + \int_0^{\infty}d\ome' \frac{\bra \Psi_\ome^+|V|\ome'\ket \bra \ome'| }{\ome-\ome'-i\eps} \bigg) \nonumber \\
   & &= |\ome\ket\bra \ome| \nonumber \\
   & &+ |\ome\ket  \bigg( \bra \Psi_\ome^+|1\ket\bra 1| + \int_0^{\infty}d\ome' \frac{\bra \Psi_\ome^+|V|\ome'\ket \bra \ome'| }{\ome-\ome'-i\eps} \bigg) \nonumber \\
   & &+ \bigg(  |1\ket\bra 1|\Psi_\ome^+\ket + \int_0^{\infty}d\ome' \frac{|\ome'\ket\bra\ome'|V|\Psi_\ome^+\ket}{\ome-\ome'+ i \eps} \bigg)\bra \ome| \nonumber \\
   & &+ \bigg(  |1\ket\bra 1|\Psi_\ome^+\ket + \int_0^{\infty}d\ome' \frac{|\ome'\ket\bra\ome'|V|\Psi_\ome^+\ket}{\ome-\ome'+ i \eps} \bigg)
   \bigg( \bra \Psi_\ome^+|1\ket\bra 1| + \int_0^{\infty}d\ome' \frac{\bra \Psi_\ome^+|V|\ome'\ket \bra \ome'| }{\ome-\ome'-i\eps} \bigg), \EQN{a3ex}
\eeqa
  we can rewrite the last term in \Ep{a3ex}
  \beqa
& &\bigg(  |1\ket\bra 1|\Psi_\ome^+\ket + \int_0^{\infty}d\ome' \frac{|\ome'\ket\bra\ome'|V|\Psi_\ome^+\ket}{\ome-\ome'+ i \eps} \bigg)
   \bigg( \bra \Psi_\ome^+|1\ket\bra 1| + \int_0^{\infty}d\ome' \frac{\bra \Psi_\ome^+|V|\ome'\ket \bra \ome'| }{\ome-\ome'-i\eps} \bigg) \nonumber \\
  & & =\bra \Psi_\ome^+|1\ket\bra 1|\Psi_\ome^+\ket  \bigg(  |1\ket + \int_0^{\infty}d\ome' \frac{|\ome'\ket}{\ome-\ome'+ i \eps} \frac{\bra\ome'|V|\Psi_\ome^+\ket}{\bra 1|\Psi_\ome^+\ket}   \bigg) \nonumber \\
  & &\times \bigg( \bra 1| + \int_0^{\infty}d\ome' \frac{ \bra \ome'| }{\ome-\ome'-i\eps}  \frac{\bra \Psi_\ome^+|V|\ome'\ket}{\bra \Psi_\ome^+|1\ket}  \bigg) \nonumber \\
  & &= (\bra \Psi_\ome^+ |1\ket-\bra 1|\Psi_\ome^+\ket) \frac{ \bra 1|\Psi_\ome^+\ket   }
 {  2\pi i \bra \ome |V|\Psi_\ome^+\ket    }  \bigg(  |1\ket + \int_0^{\infty}d\ome' \frac{|\ome'\ket}{\ome-\ome'+ i \eps} \frac{\bra\ome'|V|\Psi_\ome^+\ket}{\bra 1|\Psi_\ome^+\ket}\bigg) \nonumber \\
 & &\times  \bigg( \bra 1| + \int_0^{\infty}d\ome' \frac{ \bra \ome'| }{\ome-\ome'-i\eps} \frac{\bra \Psi_\ome^+|V|\ome'\ket}{\bra \Psi_\ome^+|1\ket}\bigg). \EQN{a4re}
   \eeqa
  From the observation
  \beqa
    & &|\ome\ket  \bigg( \bra \Psi_\ome^+|1\ket\bra 1| + \int_0^{\infty}d\ome' \frac{\bra \Psi_\ome^+|V|\ome'\ket \bra \ome'| }{\ome-\ome'-i\eps} \bigg)  \nonumber \\
    & &+ \bra \Psi_\ome^+ |1\ket \frac{ \bra 1|\Psi_\ome^+\ket   }
 {  2\pi i \bra \ome |V|\Psi_\ome^+\ket    }\bigg(  |1\ket + \int_0^{\infty}d\ome' \frac{|\ome'\ket}{\ome-\ome'+ i \eps} \frac{\bra\ome'|V|\Psi_\ome^+\ket}{\bra 1|\Psi_\ome^+\ket}\bigg) \nonumber \\
  & &\times \bigg( \bra 1| + \int_0^{\infty}d\ome' \frac{\bra \ome'| }{\ome-\ome'-i\eps}  \frac{\bra \Psi_\ome^+|V|\ome'\ket }{\bra \Psi_\ome^+|1\ket}\bigg)\nonumber \\
  & &=\bra \Psi_\ome^+ |1\ket \frac{ \bra 1|\Psi_\ome^+\ket   }
 {  2\pi i \bra \ome |V|\Psi_\ome^+\ket    }\bigg(|\ome\ket \frac{2\pi i \bra \ome |V|\Psi_\ome^+\ket }{\bra 1|\Psi_\ome^+\ket} +  |1\ket + \int_0^{\infty}d\ome' \frac{|\ome'\ket}{\ome-\ome'+ i \eps} \frac{\bra\ome'|V|\Psi_\ome^+\ket}{\bra 1|\Psi_\ome^+\ket}\bigg) \nonumber \\
  & &\times\bigg( \bra 1| + \int_0^{\infty}d\ome' \frac{ \bra \ome'| }{\ome-\ome'-i\eps} \frac{\bra \Psi_\ome^+|V|\ome'\ket}{\bra \Psi_\ome^+|1\ket} \bigg) \nonumber \\
  & &= \bra \Psi_\ome^+ |1\ket \frac{ \bra 1|\Psi_\ome^+\ket   }
 {  2\pi i \bra \ome |V|\Psi_\ome^+\ket    } \nonumber \\
  & &\times \bigg( |1\ket +  \int_0^{\infty}d\ome' \frac{|\ome'\ket}{\ome-\ome'- i \eps} \frac{\bra\ome'|V|\Psi_\ome^+\ket}{\bra 1|\Psi_\ome^+\ket} \bigg)\bigg( \bra 1| + \int_0^{\infty}d\ome' \frac{ \bra \ome'| }{\ome-\ome'-i\eps}\frac{\bra \Psi_\ome^+|V|\ome'\ket}{\bra \Psi_\ome^+|1\ket} \bigg) \EQN{a5ob}
  \eeqa
  and
   \beqa
     & &\bigg(  |1\ket\bra 1|\Psi_\ome^+\ket + \int_0^{\infty}d\ome' \frac{|\ome'\ket\bra\ome'|V|\Psi_\ome^+\ket}{\ome-\ome'+ i \eps} \bigg)\bra \ome| \nonumber \\
     & &-\bra 1|\Psi_\ome^+\ket \frac{ \bra 1|\Psi_\ome^+\ket   }
 {  2\pi i \bra \ome |V|\Psi_\ome^+\ket    }  \bigg(  |1\ket + \int_0^{\infty}d\ome' \frac{|\ome'\ket}{\ome-\ome'+ i \eps}\frac{\bra\ome'|V|\Psi_\ome^+\ket}{\bra 1|\Psi_\ome^+\ket} \bigg) \nonumber \\
 & &\times  \bigg( \bra 1| + \int_0^{\infty}d\ome' \frac{ \bra \ome'| }{\ome-\ome'-i\eps} \frac{\bra \Psi_\ome^+|V|\ome'\ket}{\bra \Psi_\ome^+|1\ket}\bigg) \nonumber \\
 & &= -\bra 1|\Psi_\ome^+\ket \frac{ \bra 1|\Psi_\ome^+\ket   }
 {  2\pi i \bra \ome |V|\Psi_\ome^+\ket    }
   \bigg(  |1\ket + \int_0^{\infty}d\ome' \frac{|\ome'\ket}{\ome-\ome'+ i \eps} \frac{\bra\ome'|V|\Psi_\ome^+\ket}{\bra 1|\Psi_\ome^+\ket}\bigg) \nonumber \\
   & &\times \bigg( - \frac{2\pi i \bra \Psi_\ome^+|V|\ome \ket}{ \bra \Psi_\ome^+|1\ket} \bra \ome|+ \bra 1| + \int_0^{\infty}d\ome' \frac{ \bra \ome'| }{\ome-\ome'-i\eps} \frac{\bra \Psi_\ome^+|V|\ome'\ket}{\bra \Psi_\ome^+|1\ket} \bigg) \nonumber \\
   & &=  -\bra 1|\Psi_\ome^+\ket \frac{ \bra 1|\Psi_\ome^+\ket   }
 {  2\pi i \bra \ome |V|\Psi_\ome^+\ket    }  \nonumber \\
 & &\times  \bigg(  |1\ket + \int_0^{\infty}d\ome' \frac{|\ome'\ket}{\ome-\ome'+ i \eps} \frac{\bra\ome'|V|\Psi_\ome^+\ket}{\bra 1|\Psi_\ome^+\ket} \bigg) \bigg( \bra 1| + \int_0^{\infty}d\ome' \frac{ \bra \ome'| }{\ome-\ome'+i\eps} \frac{\bra \Psi_\ome^+|V|\ome'\ket}{\bra \Psi_\ome^+|1\ket}\bigg), \EQN{a6Fv}
   \eeqa
    and collecting the terms (in \Ep{a6Fv} the relation \Ep{a2F} was used), we obtain \Ep{20Sep}.

\section{Completeness and orthogonality of $(\,)^{T_\pm}$} \label{App2}

 In this appendix we construct $(\,)^{T_\pm}$ operators which satisfy \Ep{56Tcomp} and \Ep{57orth}.

 For the case of $e[ \Theta B_{z} f ]^+ $, we first decompose test function $e$ which is outside the bracket
 $[\,]^+$
 \beqa
 & &\bigg(  e [\Theta B_{z} f ]^+ \bigg)^{T_-} = \bigg(  (e^+ + e^-) [\Theta B_{z} f ]^+ \bigg)^{T_-}
 = \bigg(  e^+  [\Theta B_{z} f ]^+ \bigg)^{T_-}+ \bigg( e^- [\Theta B_{z} f ]^+ \bigg)^{T_-} \nonumber \\
 & &= \bigg( e^- [\Theta B_{z} f ]^+ \bigg)^{T_-} \EQN{b1gvf}
  \eeqa
   In \Ep{b1gvf} we set
\beqa
     \bigg(  e^+  [\Theta B_{z} f ]^+ \bigg)^{T_-} \equiv 0
    \eeqa
     since it has no $H_-$ parts. When a test function $H^-$ part are multiplied with $[\,]^-$ part which might contain non-test function inside $[\,]^-$, we set that taking $( )^{T_-}$ does not change the term. This argument will be used throughout this appendix.  This also means we set
      \beqa
        \bigg(  e^+  [\Theta B_{z} f ]^+ \bigg)^{T_+} \equiv e^+  [\Theta B_{z} f ]^+.
      \eeqa 
      Remaining term becomes
     \beqa
     \bigg( e^- [\Theta B_{z} f ]^+ \bigg)^{T_-} = \bigg( e^- (\Theta B_{z} f - [\Theta B_{z} f ]^- )\bigg)^{T_-}
     = \Theta B_{z} [e^- f]^- - e^-[ \Theta B_{z} f ]^- \EQN{68gvf}
     \eeqa
     where we set
     \beqa
      \bigg( e^-[\Theta B_{z} f ]^- \bigg)^{T_-} \equiv e^-[ \Theta B_{z} f ]^-.
     \eeqa
   This also means that we set
    \beqa
       \bigg( e^-[\Theta B_{z} f ]^- \bigg)^{T_+} \equiv 0.
    \eeqa

 So we obtain here
 \beqa
  & &\bigg( e [\Theta B_{z} f ]^+ \bigg)^{T_-} = \bigg( e^- (\Theta B_{z} f - [\Theta B_{z} f ]^- )\bigg)^{T_-}
     = \Theta B_{z} [e^- f]^- - e^-[ \Theta B_{z} f ]^-,  \EQN{b7} \\
  & &   \bigg( e [\Theta B_{z} f ]^+ \bigg)^{T_+} =e^+ [B_z f]^+ + B_z [e^- f]^+. \EQN{b8}
\eeqa
 With these $\big ( e[ \Theta B_{z} f ]^+  \big)^{T\pm}$ satisfy \Ep{56Tcomp} and \Ep{57orth}. For $\big ( e[ \Theta B^c_{z} g ]^+  \big)^{T\pm}$ we can proceed in same way, only $B_z$ and $f$ replaced with $B^c_z$ and $g$.

 Last, we consider the case of
     \beqa
    \bigg( [\Theta B_{z} f ]^+\, e \, [\Theta B_{z}^{c} g ]^+ \bigg)^{T_-}. \EQN{71feg}
      \eeqa

 By decomposing $e$ as $H_+$ and $H_-$ parts, we have
      \beqa
      \bigg( [\Theta B_{z} f ]^+\, e \, [\Theta B_{z}^{c} g ]^+ \bigg)^{T_-}
      =  \bigg( [\Theta B_{z} f ]^+\, (e^- + e^+) \, [\Theta B_{z}^{c} g ]^+ \bigg)^{T_-}
      = \bigg( [\Theta B_{z} f ]^+\, e^-  \, [\Theta B_{z}^{c} g ]^+ \bigg)^{T_-}. \nonumber \\
      \eeqa
      Here we set
      \beqa
        & &\bigg( [\Theta B_{z} f ]^+\, e^+ \, [\Theta B_{z}^{c} g ]^+ \bigg)^{T_-} \equiv 0, \\
        & &\bigg( [\Theta B_{z} f ]^+\, e^+ \, [\Theta B_{z}^{c} g ]^+ \bigg)^{T_+} \equiv [\Theta B_{z} f ]^+\, e^+ \, [\Theta B_{z}^{c} g ]^+.
      \eeqa
      
      Rewriting $[\,\,]^+$ parts,
      \beqa
      & &\bigg( [\Theta B_{z} f ]^+\, e^-  \, [\Theta B_{z}^{c} g ]^+ \bigg)^{T_-}
      =  \bigg((\Theta B_{z} f- [\Theta B_{z} f ]^-) \, e^-  \,(\Theta B_{z}^{c} g -[\Theta B_{z}^{c} g ]^-) \bigg)^{T_-} \nonumber \\
      & &= \Theta B_{z} B_{z}^c \, [f e^- g]^- - \bigg( \Theta B_{z} f\,e^- [\Theta B_{z}^{c} g ]^- \bigg)^{T_-}
      - \bigg(  [\Theta B_{z} f ]^- e^-\Theta B_{z}^{c} g \bigg)^{T_-} + [\Theta B_{z} f ]^- e^- [\Theta B_{z}^{c} g ]^- \nonumber \\
      & &= \Theta B_{z}B_{z}^c  \, [f e^- g]^- - \Theta B_{z}\bigg(  f e^-[\Theta B_{z}^{c} g ]^- \bigg)^{T_-}
      - \Theta B_{z}^{c} \bigg(  [\Theta B_{z} f ]^- e^-  g \bigg)^{T_-}  \nonumber \\
      & &+ [\Theta B_{z} f ]^- e^- [\Theta B_{z}^{c} g ]^-. \EQN{b13}
      \eeqa
       Here,
        \beqa
        & &\bigg( [\Theta B_{z} f ]^- e^- [\Theta B_{z}^{c} g ]^- \bigg)^{T_-} \equiv  [\Theta B_{z} f ]^- e^- [\Theta B_{z}^{c} g ]^- ,\\
        & &  \bigg( [\Theta B_{z} f ]^- e^- [\Theta B_{z}^{c} g ]^- \bigg)^{T_+} \equiv 0.
        \eeqa
      
      In \Ep{b13},
      \beqa
      & &\bigg(  f e^-[\Theta B_{z}^{c} g ]^- \bigg)^{T_-} =  \bigg(  [f e^-]^+ [\Theta B_{z}^{c} g ]^- \bigg)^{T_-} +  [f e^-]^- [\Theta B_{z}^{c} g ]^- \nonumber \\
      & &=  \bigg(  [f e^-]^+ (\Theta B_{z}^{c} g-[\Theta B_{z}^{c} g ]^+) \bigg)^{T_-} +  f^- [\Theta B_{z}^{c} g ]^-
       \nonumber \\
      & &= \Theta B_{z}^{c} [ [f e^-]^+ g]^- + [f e^-]^- [\Theta B_{z}^{c} g ]^- \EQN{74fvg}
      \eeqa
      and similarly
      \beqa
       \bigg(  [\Theta B_{z} f ]^- e^- g \bigg)^{T_-} = [\Theta B_{z} f ]^-  [e^-g]^- +\Theta  B_{z} [ f [e^- g]^+]^-. \EQN{75vfg}
      \eeqa
      Inserting \Ep{74fvg} and \Ep{75vfg} into \Ep{b13} we obtain
\beqa
      & &\bigg( [\Theta B_{z} f ]^+\, e \, [\Theta B_{z}^{c} g ]^+ \bigg)^{T_-} \nonumber \\
      & &=\Theta B_{z} B^{c}_{z} \, [f e^- g]^-  - \Theta B_{z}B^{c}_{z}  [ [f e^-]^+ g]^- - \Theta B_{z} [f e^-]^- [\Theta B_{z}^{c} g ]^- \nonumber \\
      & &- \Theta B_{z} B^{c}_{z}[ f [e^- g]^+]^- - \Theta B_{z}^{c}[\Theta B_{z} f ]^-  [e^-g]^- + [\Theta B_{z} f ]^- e^- [\Theta B_{z}^{c} g ]^- \EQN{b18}
      \eeqa
 and
 \beqa
     & &\bigg( [\Theta B_{z} f ]^+\, e \, [\Theta B_{z}^{c} g ]^+ \bigg)^{T_+}  \nonumber \\
        & &= [\Theta B_{z} f ]^+\, e^+ \, [\Theta B_{z}^{c} g ]^+  + \Theta B_{z} B_{z}^{c} [f e^- g ]^+
        -\Theta B_{z} B_{z}^{c} \big[ [f e^-]^+  g \big]^+  \nonumber \\
        & &+ \Theta B_{z}  [f e^-]^+ [\Theta B_{z}^{c} g ]^+  - \Theta B_{z} B_{z}^{c} \big[ [e^- g]^+  f \big]^+
        + \Theta B_{z}^{c} [ e^- g]^+ [\Theta B_{z} f ]^+. \EQN{b19}
 \eeqa
 With \Ep{b18} and \Ep{b19}, for $\bigg( [\Theta B_{z} f ]^+\, e \, [\Theta B_{z}^{c} g ]^+ \bigg)^{T_\pm}$ the relation \Ep{56Tcomp} and \Ep{57orth} also hold.
 
     For  the pole $z^{c.c}$ in the upper half plane $H_+$ parts of test functions can be taken similarly.

\providecommand{\noopsort}[1]{}\providecommand{\singleletter}[1]{#1}%

\end{document}